\documentclass{jpp}
\usepackage{graphicx}

\usepackage[utf8]{inputenc}
\usepackage[T1]{fontenc}

\usepackage{amsmath, amssymb, mathrsfs, mathtools, bm}
\usepackage{color, enumerate}
\usepackage{comment}
\usepackage{orcidlink}

\usepackage{soul}

\allowdisplaybreaks

\newcommand{\im}{\mathrm{i}}
\newcommand{\bfv}{\boldsymbol{v}}

\newcommand{\bfk}{\boldsymbol{k}}
\newcommand{\bfb}{\boldsymbol{B}}
\newcommand{\bfe}{\boldsymbol{E}}
\newcommand{\bfj}{\boldsymbol{J}}
\newcommand{\bfg}{\boldsymbol{g}}
\newcommand{\bfA}{\boldsymbol{A}}
\newcommand{\bfF}{\boldsymbol{F}}
\newcommand{\bfkappa}{\boldsymbol{\kappa}}
\newcommand{\HL}{\mathscr{L}}

\newcommand{\eps}{\varepsilon}

\newcommand{\bex}{\boldsymbol{u}_1}
\newcommand{\bey}{\boldsymbol{u}_2}
\newcommand{\bez}{\boldsymbol{u}_3}

\newcommand{\prefactkappazero}{\frac{\kappa_{\parallel,0} - \kappa_{\perp,0}}{B_0^2}}

\newcommand{\Foperator}{\boldsymbol{\mathcal{F}}}

\newcommand{\Foperatorfull}{\left(\frac{k_2}{\eps}B_{02} + k_3 B_{03}\right)}

\shorttitle{Legolas: MHD spectroscopy with viscosity and Hall current}
\shortauthor{J. De Jonghe, N. Claes, and R. Keppens}

\title{\texttt{Legolas}: magnetohydrodynamic spectroscopy with viscosity and Hall current}

\author{J. De Jonghe\aff{1}
  \corresp{\email{jordi.dejonghe@kuleuven.be}} \orcidlink{0000-0003-2443-3903},
  N. Claes\aff{1} \orcidlink{0000-0002-8720-9119}
 \and R. Keppens\aff{1} \orcidlink{0000-0003-3544-2733}}

\affiliation{\aff{1}Centre for mathematical Plasma-Astrophysics, KU Leuven, 3001 Leuven, Belgium}

\begin{document}

\maketitle

\begin{abstract}
Many linear stability aspects in plasmas are heavily influenced by non-ideal effects beyond the basic ideal magnetohydrodynamics (MHD) description. Here, the extension of the modern open-source MHD spectroscopy code \texttt{Legolas} with viscosity and the Hall current is highlighted and benchmarked on a stringent set of historic and recent findings. The viscosity extension is demonstrated in a cylindrical setup featuring Taylor-Couette flow and in a viscoresistive plasma slab with a tearing mode. For the Hall extension, we show how the full eigenmode spectrum relates to the analytic dispersion relation in an infinite homogeneous medium. We quantify the Hall term influence on the resistive tearing mode in a Harris current sheet, including the effect of compressibility, which is absent in earlier studies. Furthermore, we illustrate how \texttt{Legolas} mimics the incompressible limit easily to compare to literature results. Going beyond published findings, we emphasise the importance of computing the full eigenmode spectrum, and how elements of the spectrum are modified by compressibility. These extensions allow for future stability studies with \texttt{Legolas} that are relevant to ongoing dynamo experiments, protoplanetary disks, or magnetic reconnection.
\end{abstract}

\section{Introduction}
The open-source linear magnetohydrodynamic (MHD) spectroscopy code \texttt{Legolas} was first introduced in \citet[see also \url{https://legolas.science}]{Claes2020}, and allows for computation of all linear eigenmodes and their eigenfunctions, for one-dimensionally (1D) stratified plasmas in a wide range of settings. Such non-homogeneous three-dimensional plasma states with 1D variation are very common, e.g. in plane-parallel, gravitationally stratified atmospheres, or in cylindrical setups for Taylor-Couette experiments, or in magnetic flux tubes or loops in the solar atmosphere. MHD spectroscopy is useful to determine the complete stability properties of a given force-balanced equilibrium state, and quantifies how this (in)stability is influenced by specific equilibrium ingredients, such as the magnetic pitch, or the presence of non-trivial background flows. In the original release \citep{Claes2020}, the linearised set of compressible MHD equations included the (possibly combined) effects of flow, external gravity, resistivity, anisotropic thermal conduction, and radiative cooling. The code was tested and validated against a wide variety of theoretically known plasma stability results, e.g. those from modern plasma physics textbooks focusing on MHD spectroscopy \citep{GoedbloedKeppensPoedts2019}. A typical application can be found in \citet{Claes2021}, where it was shown that especially the effect of radiative cooling in a realistic, magnetised solar atmosphere model yields MHD eigenspectra dominated by thermal instabilities and magneto-thermal overstabilities. 

Ideal MHD stability aspects have been studied extensively in the plasma physics literature, in various applications. For fusion devices such as tokamaks, a solid understanding of instabilities is required to create MHD-stable operation conditions, whilst in solar physics both stable wavemodes and instabilities are of interest to understand the observed periodicities, or the evolution of initially stable coronal loops towards destructive events such as coronal mass ejections (CMEs). The non-adiabatic effects already included in \citet{Claes2020} allow for investigating radiatively driven processes such as solar coronal rain or prominence formation. Non-ideal effects like resistivity are known to introduce new paths to instability: the well-known resistive tearing instability \citep{Furth1963} has been the subject of many studies and received renewed interest due to the role it plays in triggering magnetic reconnection events. However, physical effects that were previously omitted in \texttt{Legolas}, in particular viscosity and the Hall current, may influence growth rates or modify stability properties in a significant way.

For resistive tearing and the resulting reconnection in particular, non-linear simulations have shown that the rates at which magnetic field lines reconnect (and convert magnetic energy into kinetic or thermal energy in the process) can become much higher than resistivity can account for on its own \citep[see e.g. the review by][]{Yamada2010}. Both viscosity and Hall effects are candidates for modifying the growth rate of the resistive tearing instability. In fact, it has been known for quite some time that viscosity can act as a stabilising mechanism \citep{Coppi1966, Loureiro2013, Tenerani2015} whilst the Hall current may introduce destabilising effects, resulting in faster reconnection rates \citep{Terasawa1983, Pucci2017}. Similarly, exploration of the relationship between resistivity and viscosity on tearing by \citet{Dahlburg1983} revealed more intricacies of the growth rate as a function of the resistivity and viscosity. More recent work has focused on evaluating the influence of the Hall effect on the tearing instability in current sheets \citep{Shi2020}. We here document the extension of \texttt{Legolas} with viscous and Hall terms, and benchmark our MHD spectroscopy tool on these and some other published results. At the same time, we show how we can easily extend published findings with full spectral knowledge, or with quantifications of how incompressible and compressible regimes differ.

Whilst there are many more applications of Hall-MHD, such as the modification of the magnetorotational instability (MRI) by the Hall current in protoplanetary accretion discs \citep[see e.g. the lecture notes by][]{Lesur2021} or the influence of the Hall current on different instabilities in dynamo experiments \citep[see e.g.][]{Mishra2021}, these subjects are not treated in this current report, but are suitable to be investigated with \texttt{Legolas} in future work. In particular, \texttt{Legolas}'s multirun framework is quite apt to perform parametric studies exploring stability in function of the magnetic Prandtl number in viscoresistive setups. Such studies could also benefit from the inclusion of ambipolar diffusion, which we intend to implement in a future release of the code.

To demonstrate \texttt{Legolas}'s new capabilities, this paper is structured as follows. In section \ref{sec:framework} we present a brief overview of the compressible MHD equations implemented in \texttt{Legolas} (contrary to the incompressible equations that are often used for the tearing instability). Section \ref{sec:extensions} contains the main results of this paper, with the presentation of new diagnostic tools in section \ref{sec:def}, the viscosity module in \ref{sec:viscosity}, and the Hall module in section \ref{sec:hall}. Finally, section \ref{sec:conclusions} concludes the paper with a discussion of the preliminary results using these new \texttt{Legolas} extensions.

\section{Problem description and model equations}\label{sec:framework}
As presented in \citet{Claes2020}, but extended here with previously ignored physical effects, \texttt{Legolas} considers the full set of compressible MHD equations
\begin{align}
	\frac{\partial \rho}{\partial t} = &-\nabla \bcdot (\rho \bfv) \,, \label{eq:continuity} \\
	\rho\frac{\partial \bfv}{\partial t} = &-\nabla p - \rho \bfv \bcdot \nabla \bfv + \bfj \times \bfb + \rho\bfg + \bfF_\mathrm{visc}	\,,\label{eq:momentum} \\
	\rho\frac{\partial T}{\partial t} = &-\rho \bfv\bcdot\nabla T - (\gamma - 1)p\nabla \bcdot \bfv - (\gamma - 1)\rho\HL + (\gamma - 1)\nabla \bcdot (\bfkappa \bcdot \nabla T) \nonumber \\
	    &+ (\gamma - 1)\eta\bfj^2 + (\gamma-1) H_\mathrm{visc} \,, \label{eq:energy} \\
	\frac{\partial \bfb}{\partial t} = &\nabla \times (\bfv \times \bfb) - \nabla \times (\eta\bfj) - \nabla\times\bfj_\mathrm{Hall} \,, \label{eq:induction}
\end{align}
with variables $\rho$ density, $\bfv$ velocity, $T$ temperature, $\bfb$ magnetic field, $p = \rho T$ pressure, and $\bfj = \nabla\times\bfb$ the current density. We adopt a suitable dimensionalisation, so dimensional factors like the gas constant or the permeability of vacuum no longer appear. The adiabatic index is denoted by $\gamma$, and taken equal to $5/3$ usually. Additionally, $\bfg$ is the (external) gravitational acceleration, $\HL$ the heat loss function, defined as energy losses (optically thin radiation) minus energy gains (e.g. heating), $\bfkappa$ the thermal conduction tensor, and $\eta$ the resistivity. In this follow-up work, we focus on the additional viscous and Hall effects. The viscous force term $\bfF_\mathrm{visc}$ in the momentum equation (\ref{eq:momentum}) and the viscous heating term $H_\mathrm{visc}$ in the energy equation (\ref{eq:energy}) are introduced in section \ref{sec:viscosity}. The Hall term $\bfj_\mathrm{Hall}$ in the induction equation (\ref{eq:induction}) is detailed in section \ref{sec:hall}.

These eight non-linear partial differential equations are linearised around a one-dimensionally varying equilibrium
\begin{equation}\label{eq:equilibrium}
\begin{aligned}
    &\rho_0 = \rho_0(u_1), &&\bfv_0 = v_{02}(u_1)\,\bey + v_{03}(u_1)\,\bez \,,\\
    &T_0 = T_0(u_1), &&\bfb_0 = B_{02}(u_1)\,\bey + B_{03}(u_1)\,\bez \,,
\end{aligned}
\end{equation}
where in Cartesian coordinates $u_1$ is the $x$-coordinate, and $\bey$ and $\bez$ are the unit vectors in the then invariant $y$- and $z$-direction, respectively. In cylindrical coordinates, $u_1$ is the radial coordinate, $\bey$ is then a unit vector in the angular direction, and $\bez$ is aligned along the cylinder axis. The dependence of the background equilibrium state on the $u_1$-coordinate is considered on a bounded domain (e.g. a slab of a plane-parallel atmosphere of a given vertical extent, or a flux tube of given radius), whilst the other coordinates are unrestricted (but the $u_2$-coordinate is periodic in the cylindrical case). The linearisation introduces the perturbations $(\rho_1, \bfv_1, T_1, \bfb_1)$, which in principle are fully three-dimensionally structured, time-dependent functions. Note that the indices $0$ and $1$ refer to equilibrium and perturbed quantities, respectively.

Subsequently, after adopting a vector potential $\bfA_1$ to describe the perturbed magnetic field as $\bfb_1 = \nabla\times\bfA_1$, a 3D Fourier analysis is applied to all perturbed quantities ($\rho_1$, $\bfv_1$, $T_1$, $\bfA_1$) as
\begin{equation}\label{eq:fourier}
f_1(\mathbf{r},t) = \hat{f}_1(u_1)\,\exp\left[ \im\left( k_2 u_2 + k_3 u_3 - \omega t \right) \right],
\end{equation}
introducing the wavevector $\bfk = k_2\,\bey + k_3\,\bez$ and the frequency $\omega$ (note that in the cylindrical case, $k_2$ is an integer usually denoted by $m$, enforcing annular periodicity). In essence, this reduces the problem to a generalised eigenvalue problem
\begin{equation}\label{eq:gep}
\mathsfbi{A}\mathbf{x} = \omega\mathsfbi{B}\mathbf{x}
\end{equation}
for matrices $\mathsfbi{A}$ and $\mathsfbi{B}$, and the state vector $\mathbf{x} = (\rho_1, \bfv_1, T_1, \bfA_1)^\top$. Subsequently, it is transformed using a finite element method (FEM), where the domain is discretised using a specified number of grid points $N$ and linear combinations of basis functions are used to approximate all perturbed quantities $\hat{f}_1(u_1)$ in every subinterval. Note that this discretisation and the subsequent construction of the matrices $\mathsfbi{A}$ and $\mathsfbi{B}$ means that the number of output eigenmodes is directly related to the number of grid points $N$ \citep[for more information, see][]{Claes2020}. The resulting eigenproblem is passed to a QR solver from the LAPACK library \citep{Anderson1999}. The linearised equations of this eigenvalue problem are given in appendix \ref{app:matrix} with the inclusion of the new viscous and Hall contributions, listed in sections \ref{sec:viscosity} and \ref{sec:hall}. The \texttt{Legolas} code then returns couples of eigenvalues $\omega$ and state vectors $\mathbf{x} = (\rho_1, \bfv_1, T_1, \bfA_1)^\top$, each of which describes a fundamental linear wave of the system. Any system may have both discrete and continuous solutions, such that all of these eigenmodes in the spectrum either belong to a continuum, or correspond to a discrete solution or overtone thereof \citep[see e.g.][]{GoedbloedKeppensPoedts2019}.

For the boundaries in the $u_1$-direction we consider perfectly conducting walls, i.e.
\begin{equation}
    \bfb\bcdot\bex = 0, \qquad \bfv\bcdot\bex = 0
\end{equation}
at the edges ($\bex$ is the normal to the wall). The choice of equilibrium above guarantees that the equilibrium fields automatically satisfy these boundary conditions. For the perturbed quantities, these boundary conditions become
\begin{equation}
    v_1 = 0, \qquad k_3 A_2 - k_2 A_3 = 0,
\end{equation}
where $v_1$ is the first component of the velocity perturbation $\bfv_1 = (v_1,v_2,v_3)^\top$, and $A_2$ and $A_3$ denote components of the vector potential $\bfA_1 = (A_1, A_2, A_3)^\top$. However, instead of this second equation the more stringent condition $A_2 = A_3 = 0$ is imposed if both $k_2$ and $k_3$ are non-zero. If either wavevector component vanishes, only the corresponding $\bfA_1$-component is set to zero, i.e. if $k_2 = 0$ ($k_3 = 0$), the constraint reduces to $k_3 A_2 = 0$ ($k_2 A_3 = 0$) and only $A_2$ ($A_3$) is set to zero.

\section{Code extensions and validation results}\label{sec:extensions}
In this section, we present a couple of new features of the publicly available, open-source \texttt{Legolas} code. The first extension allows for the computation of physically relevant, derived quantities such as the perturbed magnetic field (as opposed to the auxiliary vector potential $\bfA_1$) or the entropy perturbation (as opposed to the density or temperature eigenfunction). These derived eigenfunctions are useful diagnostics, e.g. to evaluate specific eigenmode changes due to viscosity and Hall extensions. Similar to \citet{Claes2020}, these viscosity and Hall modules were tested against a fair selection of recent to historic literature results, as will be demonstrated further. However, whilst the effects of flow, gravity, resistivity, thermal conduction, and radiative cooling on full MHD spectra are relatively well-documented, the literature contains much fewer, fully reproducible tests on the linear effects of viscosity and the Hall current. Also, relevant published results were originally obtained using the incompressible MHD equations, i.e. satisfying $\nabla\bcdot\bfv = 0$ whilst typically ignoring the energy equation and/or mass conservation equation. Since the regular \texttt{Legolas} code uses the full set of compressible MHD equations, we want to be able to approximate the incompressible regime. In theory, this corresponds to the $\gamma\rightarrow\infty$ limit. In practice, we discard all terms in the energy equation's right hand side except for the term $-(\gamma-1)p_0\nabla\bcdot\bfv_1$, which relates to the finite pressure perturbation. Then, the perturbed divergence becomes $\nabla\bcdot\bfv_1 = \omega T_1 / T_0 (\gamma-1)$, whose real and imaginary parts clearly go to zero for $\gamma\rightarrow\infty$. Implementation-wise, $\gamma$ is set to a value of $10^{12}$ such that $|\nabla\bcdot\bfv_1|$ remains very small on the whole domain, usually yielding values smaller than $10^{-12}$ in the region of the spectrum near the origin. The value of $|\nabla\bcdot\bfv_1|$ is indeed observed to increase for increasingly large eigenvalues.

\subsection{Derived eigenfunctions}\label{sec:def}
There are various physical quantities of interest that can be derived from the eight eigenfunctions $(\rho_1, v_1, v_2, v_3, T_1, A_1, A_2, A_3)$ that \texttt{Legolas} computes (every eigenfunction here is now using the changed notation $f_1\equiv\hat{f}_1(u_1)$ and belonging to a specific set $(k_2, k_3, \omega)$). The most evident is the perturbed magnetic field $\bfb_1$, which is a combination of the $A_j$-eigenfunctions ($j=1,2,3$), through
\begin{equation}
\bfb_1 = \nabla\times\bfA_1 = \im\left( \frac{k_2}{\eps} A_3 - k_3 A_2 \right)\,\bex + \left( \im k_3 A_1 - A_3' \right)\,\bey + \frac{1}{\eps} \left[ (\eps A_2)' - \im k_2 A_1 \right]\,\bez \, ,
\end{equation}
where a prime denotes the derivative with respect to $u_1$ from now on. We can similarly compute its divergence $\nabla\bcdot\bfb_1$ to validate that it is numerically zero, and its curl, $\nabla\times\bfb_1$, yielding the perturbed current. Besides these magnetic-field-derived quantities, we can also determine the divergence of the velocity perturbation $\nabla\bcdot\bfv_1$, which serves as a diagnostic tool when exploiting the incompressible approximation, and the perturbed vorticity $\nabla\times\bfv_1$. Further worth mentioning is the entropy perturbation
\begin{equation}
S_1 = (p\rho^{-\gamma})_1 = \rho_0^{1-\gamma} T_1 + (1-\gamma) \rho_0^{-\gamma} T_0 \rho_1,
\end{equation}
where we used the ideal gas law $p = \rho T$ to write the entropy in terms of density and temperature.

In addition, in the presence of an equilibrium magnetic field $\bfb_0$, all perturbed vector quantities ($\bfb_1$, $\nabla\times\bfb_1$, $\bfv_1$, $\nabla\times\bfv_1$) can be expressed in a reference frame consisting of a component along the equilibrium magnetic field and two perpendicular components. This is of interest to verify or determine the (theoretically expected) polarisations of specific eigenmodes. Note that the unit vector $\bex$ is always perpendicular to the equilibrium magnetic field due to the chosen equilibrium form (\ref{eq:equilibrium}).

\subsection{Viscosity}\label{sec:viscosity}
In MHD, viscosity appears as a force term $\bfF_\mathrm{visc}$ in the momentum equation (\ref{eq:momentum}). In its most general form, $\bfF_\mathrm{visc}$ can be written as $\bfF_\mathrm{visc} = -\nabla\bcdot\boldsymbol{\upi}$, where $\boldsymbol{\upi}$ denotes the viscous stress tensor \citep{Braginskii1965}. However, as shown by \citet{Erdelyi1995}, where the authors used the full viscous stress tensor, only the shear viscosity contributes to resonant absorption, and the compressive and perpendicular components have negligible effects. Hence, for a constant dynamic viscosity $\mu$ the viscous force is in good approximation equal to \citep[see e.g.][]{GoedbloedKeppensPoedts2019}
\begin{equation}
\bfF_\mathrm{visc} = \mu\left[\nabla^2\bfv + \frac{1}{3}\nabla(\nabla \bcdot \bfv)\right].
\end{equation}
The linearisation of this expression is implemented in \texttt{Legolas}. Sometimes it is assumed that the kinematic viscosity $\nu=\mu/\rho$ is constant rather than the dynamic viscosity. This introduces additional terms in the linearisation, which are not implemented in \texttt{Legolas}. Note though that in a Cartesian setup with constant $\rho_0$ and $\bfv_0$ the additional terms introduced by assuming a constant kinematic viscosity, rather than a constant dynamic viscosity, vanish.

In addition to a contribution in the momentum equation, viscosity also adds a viscous heating term $(\gamma-1) H_\mathrm{visc}$ to the energy equation (\ref{eq:energy}). The source term $H_\mathrm{visc}$ is given by $H_\mathrm{visc} = -(\boldsymbol{\upi}\bcdot\nabla)\bcdot\bfv$ and is approximately \citep[see e.g.][]{GoedbloedKeppensPoedts2019}
\begin{equation}
H_\mathrm{visc} \approx \mu\left|\nabla \bfv \right|^2.
\end{equation}
The linearisation of this approximation in \texttt{Legolas} assumes the Frobenius norm, resulting in the linearised term
\begin{equation}
H_{\mathrm{visc},1} = 2\mu \sum\limits_{i=1}^3 \sum\limits_{j=1}^3 (\nabla\bfv_0)_{ij} (\nabla\bfv_1)_{ij}.
\end{equation}
Note that this contribution vanishes if the equilibrium flow is constant or zero, as is the case for the setup of section \ref{sec:viscoresistive} whilst it introduces two non-zero terms for the setup in section \ref{sec:taylorcouette}. However, since both setups employ the incompressible approximation, which eliminates the energy equation, this term is not represented in either test case. Note further that the background equilibrium flow $\bfv_0$ is always adopted as a stationary, Eulerian flow, much like one normally computes eigenspectra for an ideal MHD equilibrium with time-independent $\bfb_0$, even in the presence of a finite resistivity.The viscous terms are thus omitted in the equilibrium equations.

The inclusion of viscosity also imposes additional no slip boundary conditions at a rigid wall. In essence, this implies that the total plasma velocity at the boundary equals the wall's velocity. Implementation-wise, we impose that the velocity perturbation $\bfv_1$ at the boundary is exactly zero,
\begin{equation}
v_1 = v_2 = v_3 = 0.
\end{equation}
As a consequence of the no slip boundary condition, a non-zero equilibrium velocity at a boundary then simulates a boundary moving at that constant speed. We use this to study a viscous hydrodynamic Taylor-Couette flow below, which serves as a test case for cylindrical geometry by comparing to results of \citet{Gebhardt1993}. Additionally, the new viscosity module is tested in an MHD setup using the results from \citet{Dahlburg1983}, where the authors study the tearing mode in a viscoresistive plasma slab (i.e. in Cartesian geometry).

\subsubsection{Viscous eigenspectra for Taylor-Couette flow}\label{sec:taylorcouette}
When considering a viscous fluid confined between two concentric cylinders that both rotate with constant angular velocity, the flow established under no slip boundary conditions is called Taylor-Couette flow. A hydrodynamic equilibrium of this form, studied spectroscopically under incompressible conditions in \citet{Gebhardt1993}, is given by a uniform density $\rho_0$, and temperature and velocity profiles
\begin{equation}\label{eq:taylorcouette}
T_0(r) = \frac{1}{2} \left( A^2 r^2 + 4AB \log(r) - \frac{B^2}{r^2} \right) + C,\quad \bfv_0(r) = \left( Ar + \frac{B}{r} \right)\,\bey,
\end{equation}
where $C$ is an arbitrary constant to guarantee that $T_0$ is positive everywhere and
\begin{equation}
A = \frac{\beta - \left( \frac{R_1}{R_2} \right)^2 \alpha}{1 - \left( \frac{R_1}{R_2} \right)^2},\qquad B = -\frac{R_1^2 (\beta - \alpha)}{1 - \left( \frac{R_1}{R_2} \right)^2},
\end{equation}
with $\alpha$ ($\beta$) the angular speed of the inner (outer) cylinder at radius $R_1$ ($R_2$). Since this is a hydrodynamic test case, there is no equilibrium magnetic field $\bfb_0$.

\begin{figure}
\centering
\includegraphics[width=\textwidth]{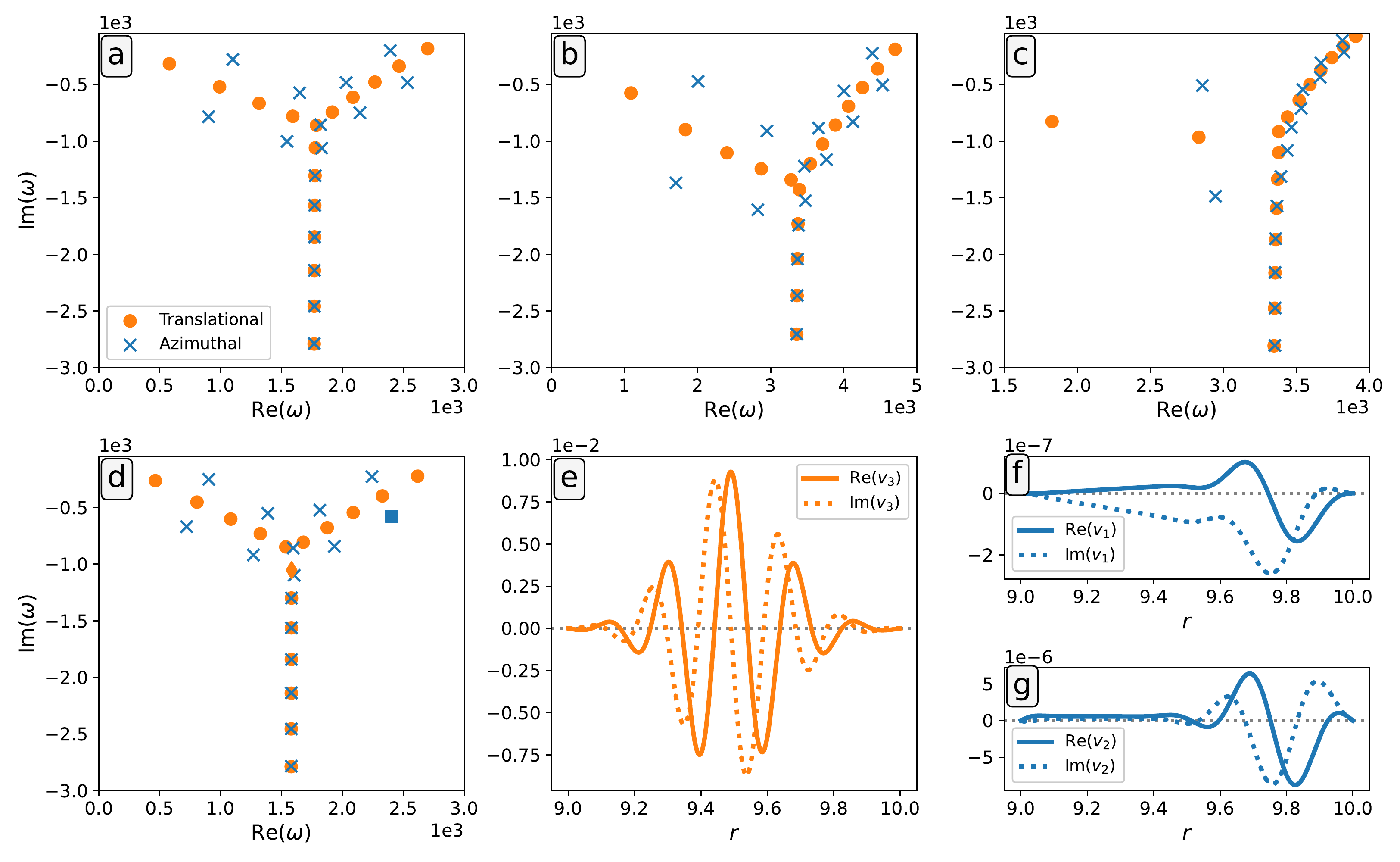}
\caption{Parts of the incompressible ($\gamma\rightarrow\infty$) Taylor-Couette spectrum with an inner cylinder at rest ($\alpha = 0$) for $k_3 = 0$, $\rho_0 = 1$, $\mu = 1$, and different parameter choices: (a) $k_2 = 1$, $R_1 = 7/3$, $R_2 = 10/3$, $\beta = 3\times 10^3$, (b) $k_2 = 2$, $R_1 = 1$, $R_2 = 2$, $\beta = 2.5\times 10^3$, (c) $k_2 = 2$, $R_1 = 0.25$, $R_2 = 1.25$, $\beta = 2\times 10^3$, and (d) $k_2 = 3$, $R_1 = 9$, $R_2 = 10$, $\beta = 10^3$. The modes represented by a dot (or $\blacklozenge$ in (d)) are translational modes with $v_1$ and $v_2$ numerically zero whilst the crosses (and $\blacksquare$ in (d)) represent azimuthal modes with non-zero $v_1$ or $v_2$ components. (e) The $v_3$ eigenfunction of the translational (d)-eigenvalue $\omega = 1583.50-1053.70\,\im$ ($\blacklozenge$). (f) and (g) show the $v_1$ and $v_2$ eigenfunction, respectively, of the azimuthal (d)-eigenvalue $\omega = 2406.82-579.55\,\im$ ($\blacksquare$). Solid lines represent real parts, dotted lines imaginary parts. All runs were performed at $251$ grid points.}
\label{fig:gebhardt-tc3}
\end{figure}

Using the incompressible approximation in \texttt{Legolas}, four representative eigenspectra from figure 3 in \citet{Gebhardt1993} are recovered in figure \ref{fig:gebhardt-tc3}a-d. These spectra feature two different types of modes, as described in \citet{Gebhardt1993}, namely translational modes with $v_1 = 0 = v_2$ (with only a non-trivial $v_3(u_1)=v_z(r)$ variation) and ``azimuthal'' modes with non-zero $v_1$ and $v_2$ eigenfunctions and $v_3 = 0$. Admittedly, we recover the azimuthal modes but they do have a non-vanishing $v_3$ eigenfunction in the test cases with \texttt{Legolas}'s incompressible approximation. However, we find that the spectra in figure \ref{fig:gebhardt-tc3}a and d are almost identical in the compressible case (the spectrum is more heavily influenced by compressibility for smaller radius-to-thickness ratios), and the $v_3$ eigenfunction indeed vanishes in the compressible setup. Hence, the incompressible approximation results in slightly spurious $v_3$ eigenfunctions for this case, presumably due to the vanishing of the $k_3v_3$ term in $\nabla\bcdot\bfv$ because $k_3$ is zero. The $v_3$ eigenfunction is shown for a translational mode in figure \ref{fig:gebhardt-tc3}e, and the $v_2$ and $v_3$ eigenfunctions of an azimuthal mode are shown in figures \ref{fig:gebhardt-tc3}f and \ref{fig:gebhardt-tc3}g, respectively. The corresponding modes are marked in figure \ref{fig:gebhardt-tc3}d. These eigenmodes in \texttt{Legolas} ($\omega_\mathrm{L}$) are consistent with those reported by \citet{Gebhardt1993} ($\omega_\mathrm{G}$), namely $\omega_\mathrm{L} = 1583.50-1053.70\,\im$ and $\omega_\mathrm{L} = 2406.82-579.55\,\im$ compared to $\omega_\mathrm{G} = 1583.56-1053.83\,\im$ and $\omega_\mathrm{G} = 2406.81-579.54\,\im$. The \texttt{Legolas} eigenfunctions match their eigenfunctions up to a complex factor (this represents the freedom to choose a reference amplitude and phase in a linear eigenvalue problem) when comparing to their figures $6$b and $8$b.

\begin{figure}
\centering
\includegraphics[width=\textwidth]{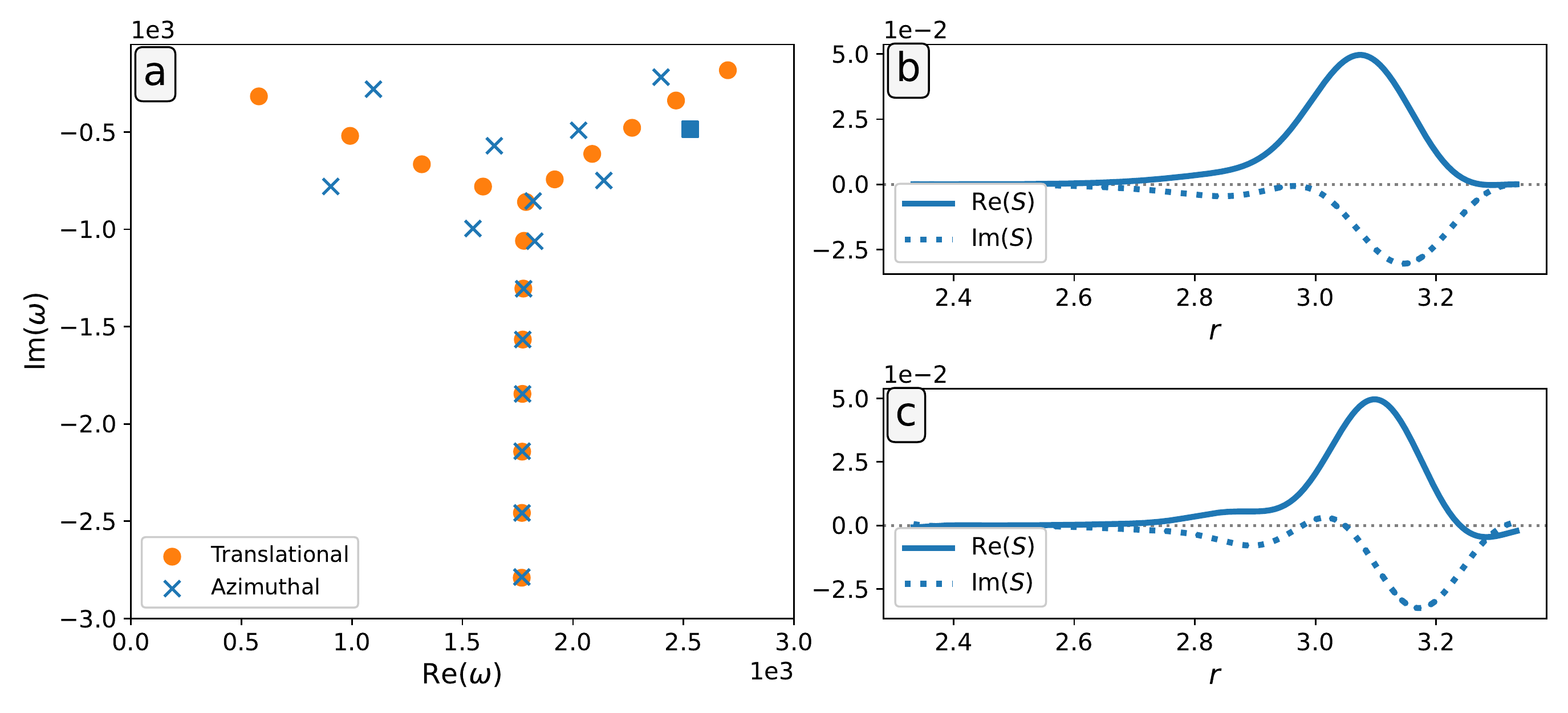}
\caption{(a) Part of the compressible Taylor-Couette spectrum (\ref{eq:taylorcouette}), with parameters $k_2 = 1$, $k_3 = 0$, $R_1 = 7/3$, $R_2 = 10/3$, $\beta = 3\times 10^3$, $\rho_0 = 1$, and $\mu = 1$. Dots represent translational modes, crosses (and $\blacksquare$) are azimuthal modes. (b) Entropy perturbation $S$ of the azimuthal mode $\omega = 2529.12 - 485.63\,\im$ ($\blacksquare$) without the inclusion of viscous heating. (c) Entropy perturbation $S$ of the azimuthal mode $\omega = 2529.66 - 485.80\,\im$ ($\blacksquare$) with the influence of viscous heating. Solid lines represent real parts, dotted lines imaginary parts. Both runs were performed at $251$ grid points.}
\label{fig:entropy}
\end{figure}

Taking the analysis of Taylor-Couette flow one step further using the fully compressible functionality of the code, we take a look at the entropy perturbation in the compressible spectrum, motivated by the observation that the azimuthal modes have a non-zero entropy perturbation whilst the translational modes have no entropy variation. In particular, we compare the entropy perturbation with and without the inclusion of viscous heating in the energy equation. Here, we use the setup of figure \ref{fig:gebhardt-tc3}a again, with parameters $k_2 = 1$, $k_3 = 0$, $R_1 = 7/3$, $R_2 = 10/3$, $\beta = 3\times 10^3$, $\rho_0 = 1$, and $\mu = 1$, but without incompressible approximation, i.e. $\gamma = 5/3$. The compressible spectrum (without viscous heating) is shown in figure \ref{fig:entropy}a. It is extremely similar to the corresponding incompressible spectrum in figure \ref{fig:gebhardt-tc3}a and is hardly influenced by viscous heating. The entropy perturbations of an azimuthal mode ($\omega = 2529.12 - 485.63\,\im$ without viscous heating; $\omega = 2529.66 - 485.80\,\im$ with viscous heating; marked by $\blacksquare$ in figure \ref{fig:entropy}a) are shown in figure \ref{fig:entropy}b and c for the compressible case without and with viscous heating, respectively. The viscous heating introduces a limited but noticeable change in the entropy perturbation $S$.

\subsubsection{Viscoresistive plasma slab}\label{sec:viscoresistive}
As a magnetohydrodynamic test case, consider the incompressible, viscoresistive stability analysis of a plane-parallel plasma slab from \citet{Dahlburg1983}, with equilibrium magnetic field profile
\begin{equation}\label{eq:dahlburg}
\bfb_0 = \left( \arctan \alpha x - \frac{\alpha x}{1+\alpha^2} \right)\,\bey
\end{equation}
with parameter $\alpha$, uniform density $\rho_0$, and $T_0$ positive and satisfying the constant total pressure condition $\partial(\rho_0 T_0 + \frac{1}{2}\bfb_0^2)/\partial x = 0$. Note that the field is not force-free, and induces a current
\begin{equation}
    \bfj_0 = \alpha \left( \frac{1}{1+\alpha^2x^2} - \frac{1}{1+\alpha^2} \right)\,\bez,
\end{equation}
which vanishes at $x=\pm 1$, where we introduce perfectly conducting walls with a no slip boundary condition. The simultaneous inclusion of resistivity and viscosity in the linear stability analysis leads to different tearing mode regimes, based on the resistivity $\eta$ and dynamic viscosity $\mu$ coefficients. The formulation in \citet{Dahlburg1983} actually uses the kinematic viscosity $\nu = \mu/\rho$, but since they assume a uniform density and no equilibrium flow, our constant $\mu$ formulation is equivalent. The relation between the resistivity $\eta$ and the kinematic viscosity $\nu$ is often expressed in terms of the magnetic Prandtl number $\mathrm{Pm} = \nu/\eta = \mu/\rho_0\eta$. In the remainder of this section, $\mathrm{Pm}$ will vary between $10^{-4}$ and $10^4$. Note that $\rho_0 = 1$ in all examples in this section, such that the Prandtl number reduces to $\mathrm{Pm} = \mu/\eta$.

In \citet{Dahlburg1983}, the authors give numerical values for the purely unstable tearing eigenmode and show the $v_{1x}$ and $B_{1x}$ eigenfunctions of the tearing mode for a few different values of $\eta$ and $\mu$ as well as the evolution of the tearing mode growth rate as a function of $\eta$, $\mu$, and the parallel wavenumber $k_2$. Here, we reproduce these results using \texttt{Legolas}.

\begin{figure}
\centering
\includegraphics[width=\textwidth]{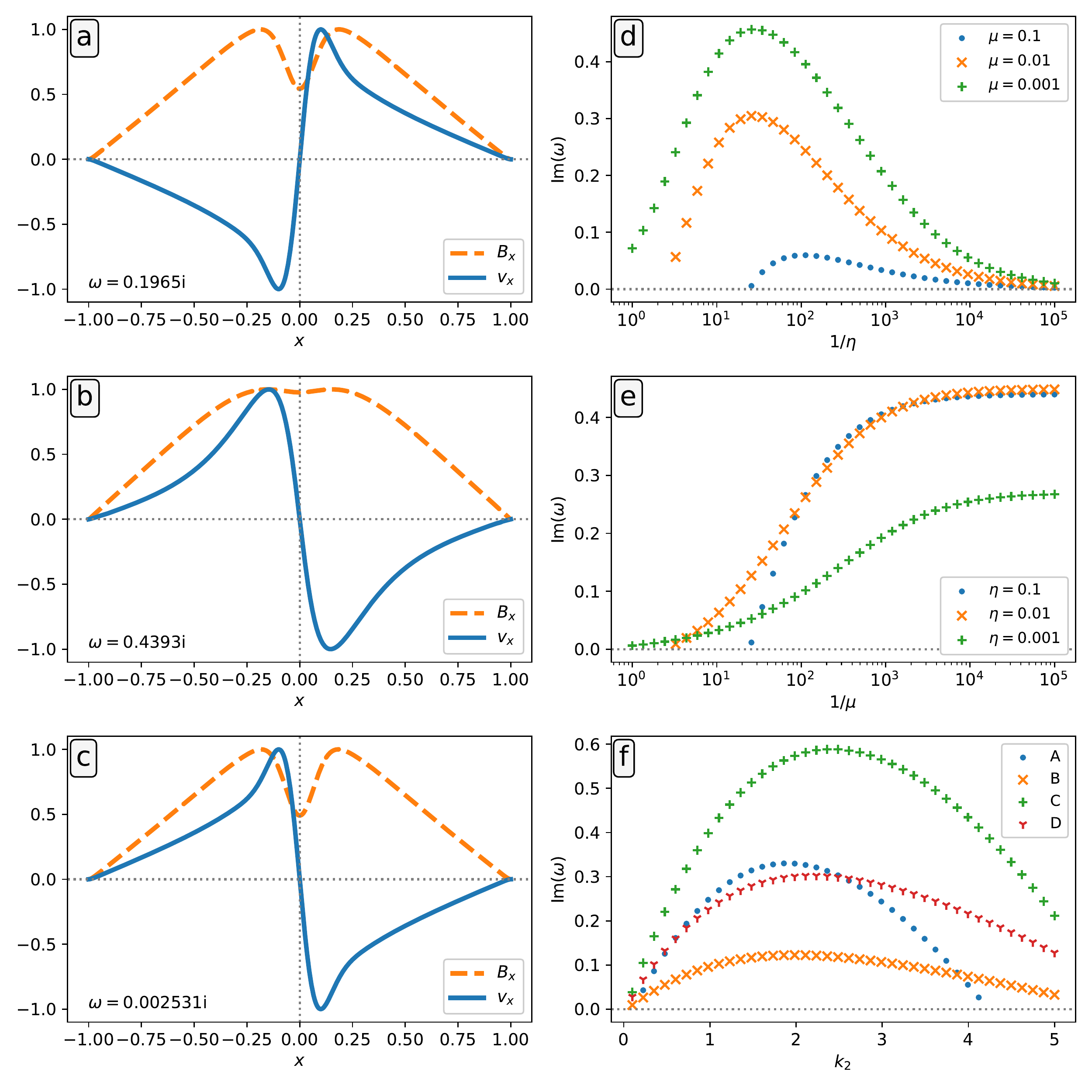}
\caption{The $v_{1x}$ and $B_{1x}$ eigenfunctions are shown for the equilibrium (\ref{eq:dahlburg}) with $\rho_0 = 1$, $\alpha = 10$, $\bfk = \bey$, and (a) $\eta = \mu = 10^{-3}$, (b) $\eta = 0.1$, $\mu = 10^{-5}$, and (c) $\eta = 10^{-5}$, $\mu = 0.1$. (d) Growth rate as a function of $\eta^{-1}$ for given values of $\mu$. (e) Growth rate as a function of $\mu^{-1}$ for given values of $\eta$. (f) Growth rate as a function of $\bfk = k_2\,\bey$ for A) $\eta = 10^{-2}$ and $\mu = 10^{-2}$; B) $\eta = 10^{-3}$ and $\mu = 10^{-2}$; C) $\eta = 10^{-2}$ and $\mu = 10^{-3}$; D) $\eta = 2\times 10^{-3}$ and $\mu = 2\times 10^{-3}$. All runs were performed at $201$ grid points.}
\label{fig:dahlburg}
\end{figure}

First, we recover the eigenvalues and eigenfunctions for three cases: (a) $\eta = \mu = 10^{-3}$, (b) $\eta = 0.1$, $\mu = 10^{-5}$, and (c) $\eta = 10^{-5}$, $\mu = 0.1$, all with $\rho_0 = 1$, $\alpha = 10$, and $\bfk = \bey$. Due to \texttt{Legolas}'s incompressible approximation, the tearing modes from \texttt{Legolas} ($\omega_\mathrm{L}$) deviate slightly from those reported in \citet{Dahlburg1983} ($\omega_\mathrm{D}$), namely (a) $\omega_\mathrm{L} = 0.1965\,\im$ compared to $\omega_\mathrm{D} = 0.19687\,\im$, (b) $\omega_\mathrm{L} = 0.4393\,\im$ compared to $\omega_\mathrm{D} = 0.4397\,\im$, and (c) $\omega_\mathrm{L} = 0.002531\,\im$ compared to $\omega_\mathrm{D} = 0.002537\,\im$. The $v_{1x}$ and $B_{1x}$ eigenfunctions, defined up to a complex factor and rescaled here for comparison to figures 2C, D, and E in \citet{Dahlburg1983}, are shown in figures \ref{fig:dahlburg}a, b, and c. The arbitrary complex factor is chosen for each case such that all shown eigenfunctions are real.

Next, we reproduce the evolution of the tearing mode growth rate (d) as a function of $\eta^{-1}$ for fixed values of $\mu$, (e) as a function of $\mu^{-1}$ for fixed values of $\eta$, and (f) as a function of $k_2$ for fixed values of $\eta$ and $\mu$. In figures \ref{fig:dahlburg}d, e, and f the tearing growth rate is shown for each configuration of parameters, obtaining the positive growth rates shown in \citet{Dahlburg1983} in their figures 6A, 7, and 9, respectively. Once again, they agree very well. Note that each marker represents a single \texttt{Legolas} run at $201$ grid points.

\begin{figure}
\centering
\includegraphics[width=\textwidth]{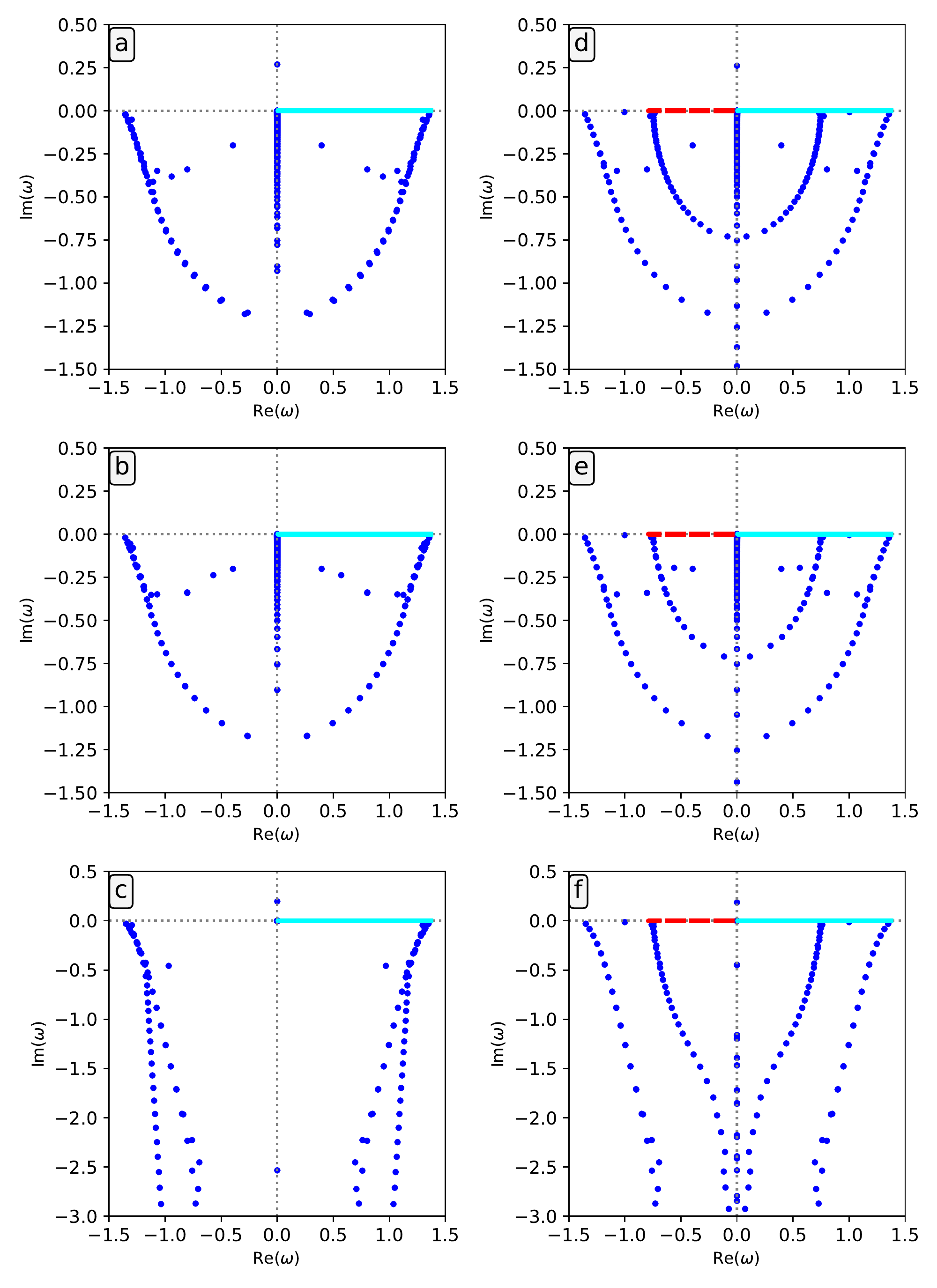}
\caption{Comparison of spectra for equilibrium profile (\ref{eq:dahlburg}) for resistive (a, d), viscous (b, e), and viscoresistive (c, f) cases. The left column (a, b, and c) represents the incompressible approximation, the right column (d, e, and f) the compressible equations. All runs use parameters $\bfk = \bey$, $\rho_0 = 1$, and $\alpha = 10$. In case of resistivity (viscosity), the parameter is $\eta = 10^{-3}$ ($\mu = 10^{-3}$). The cyan solid and red dashed lines represent the ideal MHD Alfv\'en and slow continua, respectively. Both continua are symmetric with respect to the imaginary axis, but only shown in one halfplane to avoid overlap. All runs were performed at $251$ grid points.}
\label{fig:viscoresistive}
\end{figure}

Unlike \citet{Dahlburg1983}, \texttt{Legolas} does not only compute the tearing mode, but the entire spectrum. Hence, we can also compare the purely resistive, purely viscous, and truly viscoresistive spectra for the same equilibrium profile (\ref{eq:dahlburg}). This is shown in figure \ref{fig:viscoresistive} for runs at $251$ grid points. In this figure, the left column displays the incompressible limit and the right column shows the fully compressible spectra. All spectra are supplemented with the analytical, ideal MHD slow and Alfv\'en continua, which correspond to singular solutions of the ordinary differential equation obtained through a reformulation of the ideal MHD equations in terms of the $x$-component of the Lagrangian displacement field. It can be shown for homogeneous backgrounds \citep[see e.g.][]{GoedbloedKeppensPoedts2019} that in the presence of resistivity the Alfv\'en and slow modes trace out semi-circles in the stable part of the spectrum with infinitely degenerate (collapsed) continua. For inhomogeneous resistive spectra the ideal continuum ranges will relocate to collections of discrete modes in the stable half-plane, still resembling the semi-circular curves, as seen in figures \ref{fig:viscoresistive}a and d, which will have links to extremal or edge values of the ideal continua. Figures \ref{fig:viscoresistive}b and e now show that viscosity exerts a similar influence as resistivity. Finally, figures \ref{fig:viscoresistive}c and f represent modified variants of the semi-circle-like curves in the other panels, due to the combined effects of viscosity and resistivity. Since the ideal slow and Alfv\'en continua partially overlap and are symmetric with respect to the imaginary axis, the slow continuum is only drawn in the left halfplane (red dashed line) and the Alfv\'en continuum in the right halfplane (cyan solid line). In the left column, the slow continua are eliminated by the incompressible assumption.

The first row of figure \ref{fig:viscoresistive} (panels a and d) shows the resistive slab with $\eta = 10^{-3}$. This case is well known and discussed in e.g. \citet{GoedbloedKeppensPoedts2019}. In panel a, the Alfv\'en modes form a semicircle and the slow (magnetoacoustic) modes are eliminated by the incompressible approximation. In the compressible case of panel d, the slow modes reappear as the inner semicircle. Finally, both the compressible and incompressible spectra feature a resistive tearing mode, as the only purely unstable eigenmode of this system for the chosen parameters.

The second row of figure \ref{fig:viscoresistive} (panels b and e) on the other hand shows the viscous case with $\mu = 10^{-3}$. The result looks surprisingly similar to the resistive case, with both the slow and Alfv\'en modes taking on the same semicircular shape of similar magnitude (note that the axes are scaled identically in panels a, b, d, and e). Whilst there are many minute differences with the resistive case in the first row, the key difference is the absence of a tearing mode in the viscous spectra.

Ultimately, the third row (panels c and f) shows the viscoresistive spectrum with $\eta = \mu = 10^{-3}$. Although resistivity and viscosity exert a similar influence on the slow and Alfv\'en modes when they are the only physical effect in consideration, the combination of both effects reveals new behaviour in both the incompressible (panel c) and compressible case (panel f). Whilst the slow and Alfv\'en branches still originate in the same point on the real axis, the semicircular structures are replaced by stretched-out curves along the imaginary axis. The resistive tearing mode is still present, but damped by the viscosity. Therefore, the changes are most pronounced on the stable and damped parts of the spectrum, whose physical relevance must also consider the fact that the ideal MHD equilibrium itself will evolve on a specific diffusive timescale when viscoresistive effects are active.

\subsection{Hall-MHD}\label{sec:hall}
The induction equation, given by the Maxwell-Faraday equation
\begin{equation}
\frac{\partial\bfb}{\partial t} = -\nabla\times\bfe,
\end{equation}
can be extended to include the effects of the Hall current, electron pressure, and electron inertia by expressing the electric field $\bfe$ using the (dimensionless) generalised Ohm's law
\begin{equation}\label{eq:ohm}
\bfe = -\bfv\times\bfb + \eta\bfj + \frac{\eta_\mathrm{H}}{\rho} \left( \bfj\times\bfb - \nabla p_\mathrm{e} \right) + \frac{\eta_\mathrm{e}}{\rho} \frac{\partial\bfj}{\partial t}.
\end{equation}
Here, $p_\mathrm{e}$ denotes the electron pressure and is related to $p$ through the electron fraction $f_\mathrm{e}$ as $p_\mathrm{e} = f_\mathrm{e}p$, with $f_\mathrm{e} = n_\mathrm{e} / (n_\mathrm{e}+n_\mathrm{p}) = 1/2$ for a charge-neutral electron-proton plasma. Furthermore, $\eta_\mathrm{H}$ and $\eta_\mathrm{e}$ are the normalised Hall and electron inertia coefficients,
\begin{equation}
    \eta_\mathrm{H} = \frac{m_\mathrm{i}}{e} \frac{V_\mathrm{R}}{L_\mathrm{R} B_\mathrm{R}},\qquad \eta_\mathrm{e} = \frac{m_\mathrm{e} m_\mathrm{i}}{e^2} \left( \frac{V_\mathrm{R}}{L_\mathrm{R} B_\mathrm{R}} \right)^2,
\end{equation}
respectively. Here, $e$ denotes the fundamental charge, $m_\mathrm{i}$ and $m_\mathrm{e}$ are the ion and electron mass respectively, and $V_\mathrm{R}$, $L_\mathrm{R}$, and $B_\mathrm{R}$ are the reference velocity, length, and magnetic field strength. Consequently, the electron inertia coefficient $\eta_\mathrm{e}$ is several orders of magnitude smaller than the Hall coefficient $\eta_\mathrm{H}$. Therefore, the effect of electron inertia is usually negligible. Hence, most results in the literature do not include it. Whilst this effect is implemented in \texttt{Legolas}, the reference tests that follow all set $\eta_e = 0$, so its effect is only quantified for one limit case here. It should also be pointed out that any equilibrium of the form (\ref{eq:equilibrium}) that satisfies the ideal MHD equilibrium conditions, also satisfies the Hall-MHD equilibrium conditions (neglecting electron inertia) because the Hall term in the induction equation (\ref{eq:induction}), given by
\begin{equation}
    \nabla\times\bfj_{\mathrm{Hall},0} = \nabla\times \left[ \frac{\eta_\mathrm{H}}{\rho_0} \left( \bfj_0\times\bfb_0 - \nabla p_{\mathrm{e}0} \right) \right],
\end{equation}
reduces to zero. For the first term this follows because $\bfb_0$ and $\bfj_0$ both lie in the $\bey\bez$-plane, such that their vector product is proportional to $\bex$, and since they only depend on $u_1$, this implies that $\nabla\times f(u_1)\,\bex = 0$. The second term is the curl of a gradient, which is always zero. Note that whilst there are many similarities between the resistive and Hall terms, here they differ since the resistive term does not disappear in the induction equation (\ref{eq:induction}) for an equilibrium of the form (\ref{eq:equilibrium}). As pointed out in \citet{Claes2020}, the resistive term is neglected in the equilibrium equations by assuming that the timescales on which magnetic fields decay is much larger than the timescales of resistive modes.

The Hall and electron pressure terms in the induction equation are not implemented directly in \texttt{Legolas} as written in equation (\ref{eq:ohm}). Instead, $\bfj\times\bfb$ is substituted into this expression using the momentum equation (\ref{eq:momentum}) as done in e.g. \citet{Ahedo2009} because it is observed to be more stable numerically. The result is
\begin{equation}
\bfe = -\bfv\times\bfb + \eta\bfj + \eta_\mathrm{H} \left\{ \frac{\partial\bfv}{\partial t} + \bfv\bcdot\nabla\bfv - \bfg - \frac{\mu}{\rho} \left[ \nabla^2\bfv + \frac{1}{3}\nabla(\nabla\bcdot\bfv) \right] + \frac{\nabla p_\mathrm{i}}{\rho} \right\} + \frac{\eta_\mathrm{e}}{\rho} \frac{\partial\bfj}{\partial t}. \label{eq:efield}
\end{equation}
This equation now features the ion pressure $p_i$ instead, which is related to the total pressure as $p_\mathrm{i} = (1-f_\mathrm{e})p$. Note that this expression for the electric field now has two time derivatives in the right hand side, which will then enter the induction equation. Exploiting $\bfA_1$ instead of $\bfb_1$, the linearised induction equation becomes $\partial \bfA_1/\partial t = -\bfe_1$. Hence, we linearise equation (\ref{eq:efield}), allowing for a temperature-dependent Spitzer resistivity $\eta(T)$, which gives
\begin{equation}\label{eq:E1}
\begin{aligned}
\bfe_1 = &-\bfv_1\times\bfb_0 - \bfv_0\times\left(\nabla\times\bfA_1\right) + \eta_0 \nabla\times\left(\nabla\times\bfA_1\right) + \frac{\mathrm{d}\eta}{\mathrm{d}T} T_1 \nabla\times\bfb_0 \\
&+\eta_\mathrm{H} \bigg\{ \frac{\partial\bfv_1}{\partial t} + \bfv_1\bcdot\nabla\bfv_0 + \bfv_0\bcdot\nabla\bfv_1 - \frac{\mu}{\rho_0} \left[ \nabla^2\bfv_1 + \frac{1}{3}\nabla\left(\nabla\bcdot\bfv_1\right) \right] \\
&\qquad+ \mu\frac{\rho_1}{\rho_0^2} \left[ \nabla^2\bfv_0 + \frac{1}{3}\nabla\left(\nabla\bcdot\bfv_0\right) \right] + \frac{\nabla p_{\mathrm{i}1}}{\rho_0} - \frac{\rho_1\nabla p_{\mathrm{i}0}}{\rho_0^2} \bigg\} \\
&+ \frac{\eta_\mathrm{e}}{\rho_0}\frac{\partial}{\partial t} \left[ \nabla\times(\nabla\times\bfA_1) \right] - \eta_\mathrm{e}\frac{\rho_1}{\rho_0^2} \frac{\partial}{\partial t} \left( \nabla\times\bfb_0 \right).
\end{aligned}
\end{equation}
This expression can be simplified by observing that the term $\nabla p_{\mathrm{i}1}/\rho_0$ can be written as
\begin{equation}\label{eq:pi-grad}
\frac{\nabla p_{\mathrm{i}1}}{\rho_0} = \nabla\left( \frac{p_{\mathrm{i}1}}{\rho_0} \right) + \frac{\nabla\rho_0}{\rho_0^2} p_{\mathrm{i}1}.
\end{equation}
Hence, this term is a pure gradient if $\rho_0$ is uniform. Since the electric field is only defined up to a gradient, we can redefine it as $\tilde{\bfe}_1 = \bfe_1 - \eta_\mathrm{H} \nabla (p_{\mathrm{i}1}/\rho_0)$ (with $\bfe_1$ expression (\ref{eq:E1})) such that  after substituting equation (\ref{eq:pi-grad}) $p_{\mathrm{i}1}$ only appears in $\tilde{\bfe}_1$ in the term $\eta_\mathrm{H} p_{\mathrm{i}1}\nabla\rho_0/\rho_0^2$, and thus only if $\rho_0$ is not uniform. The resulting induction equation $\partial \bfA_1/\partial t = -\tilde{\bfe}_1$ is implemented in the code. Note that in the generalised eigenvalue problem (\ref{eq:gep}) resulting from the Fourier analysis (\ref{eq:fourier}) the time derivatives in the Hall and electron inertia terms enter in the $\mathsfbi{B}$-matrix and break its former symmetry.

In what follows, we present a series of stringent test cases to validate our Hall-MHD linear solver.

\subsubsection{Hall-MHD waves in a homogeneous plasma slab}
For the first test case, consider a homogeneous Cartesian plasma slab confined between two perfectly conducting walls (perpendicular to the $x$-axis). The equilibrium is given by
\begin{equation}
\rho_0 = 1, \quad T_0 = 1, \quad \bfb_0 = \bez \,,
\end{equation}
and our normalisation is chosen such that $\eta_\mathrm{H} = 1$. We want to quantify Hall-MHD eigenmodes of this slab, where we can compare to the analytical result of waves for an infinite homogeneous plasma in \citet{Hameiri2005}, where they give the dimensionful dispersion relation
\begin{equation}\label{eq:hameiri}
\begin{aligned}
&\left( \frac{\omega}{kv_{\mathrm{A}0}} \right)^6 - \left( \frac{\omega}{kv_{\mathrm{A}0}} \right)^4 \left[ 1 + \frac{\gamma T_0}{v_{\mathrm{A}0}^2} + \cos^2\theta \left( 1 + \frac{(k\eta_\mathrm{H})^2}{\rho_0} \right) \right] \\
&\quad + \left( \frac{\omega}{kv_{\mathrm{A}0}} \right)^2 \cos^2\theta \left[ 1 + \frac{\gamma T_0}{v_{\mathrm{A}0}^2} \left( 2 + \frac{(k\eta_\mathrm{H})^2}{\rho_0} \right) \right] - \frac{\gamma T_0}{v_{\mathrm{A}0}^2}\ \cos^4\theta = 0.
\end{aligned}
\end{equation}
Here, $k$ is the magnitude of the wavevector $\bfk$, $\theta$ is the angle between $\bfk$ and $\bfb_0$, and $v_{\mathrm{A}0} = |\bfb_0|/\sqrt{\mu_0\rho_0}$ is the equilibrium Alfv\'en speed, where $\mu_0$ is the vacuum permeability. The inclusion of the Hall term introduces a length scale into the previously scale-independent MHD equations through the ion skin depth $d_\mathrm{i} = \eta_\mathrm{H}/\sqrt{\rho_0}$. This makes the Hall-MHD waves dispersive as seen from the above dispersion relation. To simulate an infinite medium, we need to ensure that the ratio of the equilibrium ion skin depth to the system size is sufficiently small. Hence, for the choice of $d_{\mathrm{i}0} = 1$, we solve in the interval $x\in[0,10^3]$. The exact choice of interval size is largely arbitrary, but it should be kept in mind that when we increase the interval size, we may also be forced to increase the resolution to ensure that \texttt{Legolas} picks up the Hall modes. This is due to the fact that the grid resolution can be directly linked to the dimensions of the $\mathsfbi{A}$ and $\mathsfbi{B}$ matrices in the eigenvalue problem \ref{eq:gep}, and thus also to the number of eigenvalues returned.

\begin{figure}
\centering
\includegraphics[width=\textwidth]{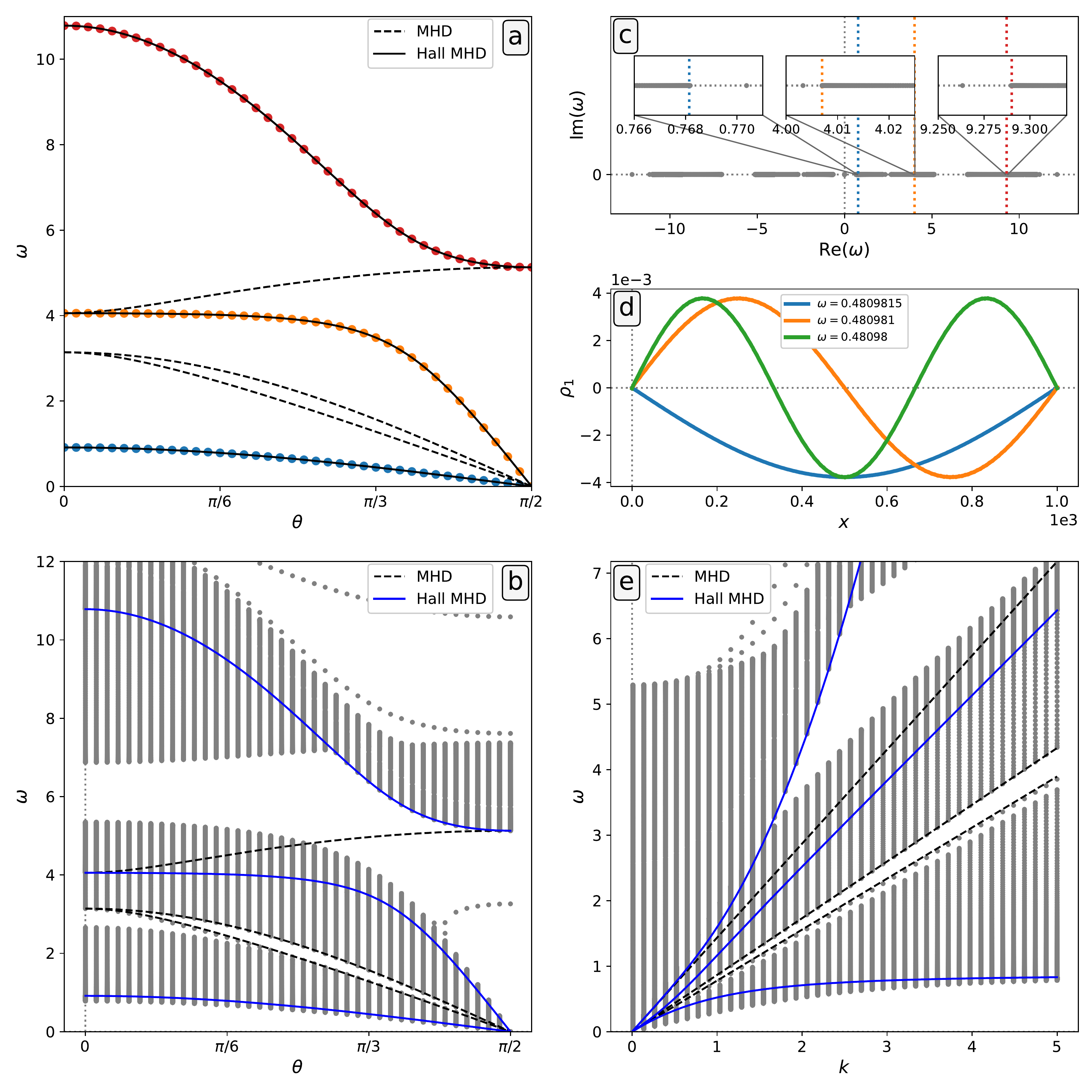}
\caption{(a) Comparison of the first mode of each sequence (dots) to the theoretical Hall prediction by \citet{Hameiri2005} (solid lines) and ideal MHD (dashed lines) as a function of the angle $\theta$ between $\bfk = \upi\,(\sin\theta\,\bey + \cos\theta\,\bez)$ and $\bfb_0 = \bez$ with $\rho_0 = 1$, $T_0 = 1$, $\eta_\mathrm{H} = 1$, and $x\in [0,10^3]$. (b) Comparison of the full spectrum to MHD and Hall-MHD solutions for the setup from (a). (The isolated branches are unresolved modes.) (c) Spectrum for an angle $\theta\approx 0.564$ with the three (positive) solutions of the dispersion relation (\ref{eq:hameiri}) as vertical (dotted) lines. (d) $\rho_1$ eigenfunctions of the first three modes of the smallest solution sequence for $\theta\approx 1.007$. (e) Comparison of the full spectrum to ideal and Hall-MHD predictions for varying wavenumber for $\bfk = k\,(\bey/2 + \sqrt{3}\,\bez/2)$, $\bfb_0 = \bez$, $\rho_0 = 1$, $T_0 = 1$, $\eta_\mathrm{H} = 1$, and $x\in [0,10^3]$. All runs were performed at $501$ grid points.}
\label{fig:adiabhomo}
\end{figure}

Since the medium in \texttt{Legolas} is bounded in the $x$-direction, each solution of the dispersion relation (\ref{eq:hameiri}) should approximate the first mode in a sequence in the spectrum, which can be verified by the number of nodes in the mode's corresponding eigenfunctions. For given angles $\theta$, the first mode of each sequence is shown in figure \ref{fig:adiabhomo}a alongside the theoretical curves and a comparison to the ideal MHD dispersion relation. A full spectrum version is also shown in figure \ref{fig:adiabhomo}b. As can be seen in equation (\ref{eq:hameiri}), in the case of perpendicular propagation the dispersion relation is not influenced by the Hall parameter ($\cos\theta = 0$). There, the highest mode reduces to the regular fast MHD mode and the lower two modes (slow and Alfv\'en) vanish, also visible in figures \ref{fig:adiabhomo}a and b.  The spectrum itself is shown for an angle $\theta \approx 0.564$ in figure \ref{fig:adiabhomo}c alongside the analytical infinite-medium solutions, each one corresponding to the start of a sequence, indicated by vertical lines. The sequences themselves, whose modes are much more tightly packed than in the ideal MHD sequences, are shown in the insets of figure \ref{fig:adiabhomo}c. The smallest sequence displays anti-Sturmian behaviour, similar to the ideal MHD slow modes, whilst the two larger sequences behave in a Sturmian way, like the ideal MHD Alfv\'en and fast modes.

Furthermore, the (real) $\rho_1$ eigenfunctions of the first three modes in the smallest sequence of the $\theta \approx 1.007$ spectrum are given in figure \ref{fig:adiabhomo}d. Contrary to the slow, Alfv\'en, and fast modes in ideal MHD, the density perturbation vanishes at the edges here. This behaviour is easily derived from the equations in appendix \ref{app:matrix} for the adiabatic homogeneous setup considered here. Neglecting equilibrium flow, resistivity, and viscosity, and applying the perfectly conducting wall boundary conditions $\tilde{v}_1 = \tilde{a}_2 = \tilde{a}_3 = 0$ (with the tildes indicating the transformed variables \ref{eq:coordtransform}) reduces the second and third components of the induction equation (\ref{eq:app-induction2}) and (\ref{eq:app-induction3}) to $\tilde{v}_2 = 0$ and $\tilde{v}_3 = 0$, respectively, for non-zero frequency. Since these equations vanish altogether for an adiabatic homogeneous setup in ideal MHD, these emerging no slip boundary conditions are naturally imposed by the Hall current. Using these newfound conditions alongside the others in the continuity equation (\ref{eq:app-continuity}), the third component of the momentum equation (\ref{eq:app-momentum3}), and the energy equation (\ref{eq:app-energy}) reduces these equations to $\omega\tilde{\rho}_1 = -\rho_0 \tilde{v}_1'$, $k_3 (\tilde{\rho}_1 T_0 + \rho_0 \tilde{T}_1) = 0$, and $\omega \tilde{T}_1 = -(\gamma-1)T_0 \tilde{v}_1'$, respectively, where we also used that $B_{02} = 0$ in our reference frame (for a different reference frame a linear combination of (\ref{eq:app-momentum2}) and (\ref{eq:app-momentum3}) gives similar results for any constant $\bfb_0$). Combining all three implies that $\tilde{\rho}_1 = \tilde{T}_1 = \tilde{v}_1' = 0$ at the wall boundaries. Since this derivation made no assumptions about the waves, this behaviour is present for all modes in the adiabatic homogeneous Hall spectrum. Note however that it only holds for oblique angles between $\bfk$ and $\bfb_0$. If $\bfk$ is parallel to $\bfb_0$ and along the $y$- or $z$-axis in a reference frame of choice, either $\tilde{a}_2 = 0$ or $\tilde{a}_3 = 0$ does not hold and therefore the other equations do not reduce in the way described above. Therefore, for parallel propagation the density perturbation is non-zero at the boundaries, just like in the ideal MHD case.

Finally, figure \ref{fig:adiabhomo}e shows the dispersion of all three sequences by comparing the full spectrum for different wavenumbers to the theoretical ideal MHD and Hall-MHD predictions (the dispersive nature of the middle sequence is more subtle). This dispersive behaviour identifies the largest sequence as the so-called whistler wave, which derives its name from the property that higher frequencies travel faster such that higher frequencies reach observers at earlier times than lower frequencies. The smallest sequence is sometimes called the ion cyclotron wave because its frequency approaches $\Omega_\mathrm{i}\cos\theta$ asymptotically for increasing wavenumber, where $\Omega_\mathrm{i} = ZeB_0/m_\mathrm{i}$ is the ion cyclotron frequency with $Z$ the charge number, $e$ the fundamental charge, and $m_\mathrm{i}$ the ion mass (in \texttt{Legolas}, the ions are protons). Note that the final (intermediate) mode in this panel, which is related to the MHD Alfv\'en wave and two-fluid A mode \citep{DeJonghe2020}, fails to capture the electron cyclotron resonance $\omega\rightarrow\Omega_\mathrm{e}\cos\theta$ ($\Omega_\mathrm{e} = eB_0/m_\mathrm{e}$, with $m_\mathrm{e}$ the electron mass) in the short wavelength (large wavenumber) limit, which is present in the two-fluid description, because $\eta_\mathrm{e}$ ($\propto m_\mathrm{e}$) was set to zero \citep{Hameiri2005}. It has been verified that the sequences indeed start at the theoretical Hall-MHD results (up to an error of $10^{-5}$ at $501$ grid points and a ratio of $10^{-3}$ of $\eta_\mathrm{H}$ to slab thickness), even though it is somewhat unclear in this image due to the large frequency range and the proximity of various sequences.

\begin{figure}
\centering
\includegraphics[width=\textwidth]{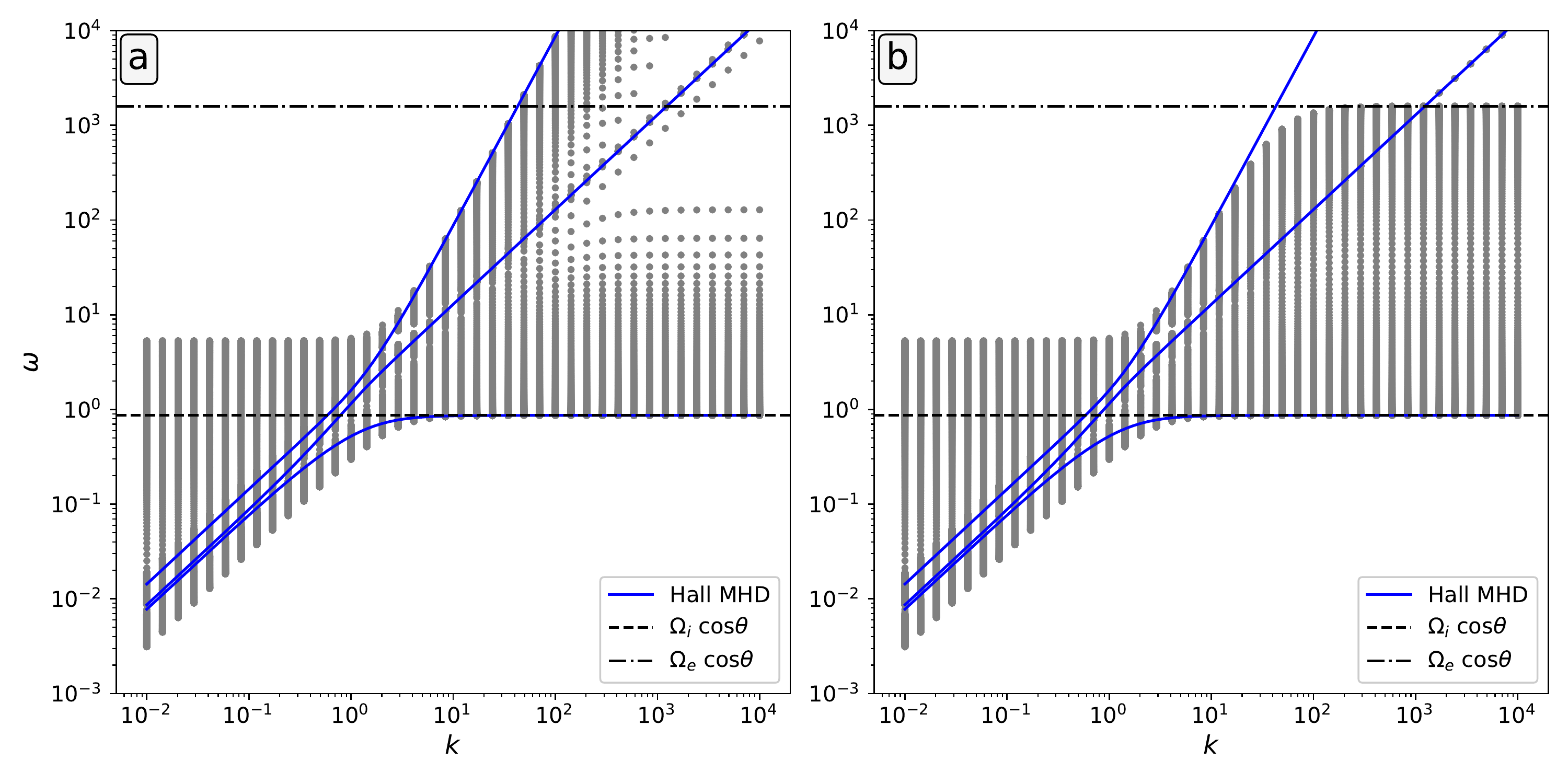}
\caption{Comparison to the theoretical Hall-MHD curves from \citet[where $\eta_\mathrm{e} = 0$]{Hameiri2005} and two-fluid resonances ($\Omega_\mathrm{e}\cos\theta$, $\Omega_\mathrm{i}\cos\theta$) of the frequency $\omega$ as a function of wavenumber $k$ in Hall-MHD (a) without electron inertia ($\eta_\mathrm{e} = 0$) and (b) with electron inertia ($\eta_\mathrm{e} \neq 0$). The setup is identical to the one used in figure \ref{fig:adiabhomo}e. All runs were performed at $501$ grid points.}
\label{fig:elecinertia}
\end{figure}

As a quick test of the electron inertia term, it is shown in figure \ref{fig:elecinertia} that the inclusion of this term ($\eta_\mathrm{e} \neq 0$) captures the behaviour near the electron cyclotron resonance for large wavenumbers.

\subsubsection{Resistive Harris sheet: tearing in Hall-MHD}\label{sec:harris}
In \citet{Shi2020}, the authors investigate the influence of the Hall current on the resistive tearing mode of a Harris current sheet. The equilibrium profile takes the form
\begin{equation}\label{equil:hall-tearing}
\rho_0 = \tilde{\rho}_0,\quad T_0 = \frac{B_0^2}{2\tilde{\rho}_0}\ \mathrm{sech}^2 \left( \frac{x}{a} \right),\quad \bfb_0 = B_0 \tanh\left( \frac{x}{a} \right)\,\bey + B_\mathrm{g}\,\bez
\end{equation}
with $\tilde{\rho}_0 = B_0 = a =1$ and $B_\mathrm{g}$ a variable guide field parameter. The included physical effects are a constant resistivity $\eta = 10^{-4}$ and a Hall current with coefficient $\eta_\mathrm{H} = 1$. As explained earlier, in the vector potential formulation in \texttt{Legolas} we include the Hall term and the electron pressure term, but in this test we ignore the electron inertia effect (i.e. $\eta_e=0$). Furthermore, \citet{Shi2020} assume incompressibility, so we also use the incompressible approximation (large $\gamma$) in \texttt{Legolas}.

\begin{figure}
\centering
\includegraphics[width=\textwidth]{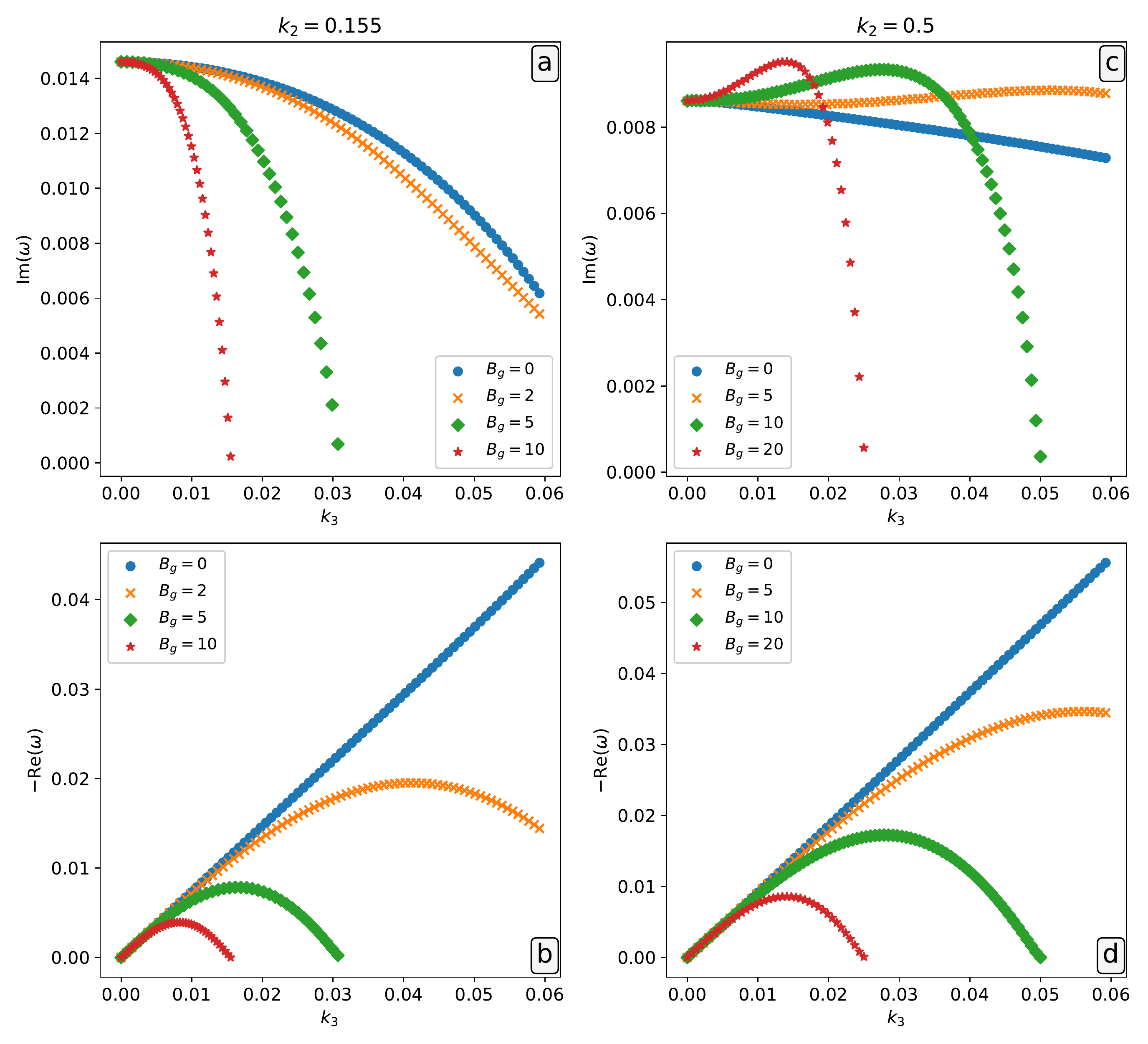}
\caption{The real (b, d) and imaginary (a, c) parts of the tearing mode in a Harris current sheet (\ref{equil:hall-tearing}) as a function of $k_3$, with $\tilde{\rho}=1$, $a=1$, $B_0=1$, $\eta = 10^{-4}$, and $\eta_\mathrm{H} = 1$ for different guide field strengths $B_\mathrm{g}$. (a) and (b) correspond to a wavevector $\bfk = 0.155\,\bey$, and (c) and (d) to $\bfk = 0.5\,\bey$. All runs were performed at $501$ grid points.}
\label{fig:harris}
\end{figure}

\citet{Shi2020} solve for the tearing mode on the interval $x\in [-15,15]$ and assume exponential decay of the perturbed quantities outside of that interval since the profile (\ref{equil:hall-tearing}) is approximately constant there for the chosen parameters. In \texttt{Legolas}, the default boundary settings are conducting wall boundary conditions at a finite distance, which may modify the linear MHD spectrum due to e.g. wall stabilisation effects. However, for $a = 1$ the interval $[-15,15]$ seems large enough such that the stabilising influence of the conducting walls is expected to be negligible. Our results are shown for two different values of $k_2$ in figure \ref{fig:harris}, recovering figure 4 in \citet{Shi2020}. In this figure, we show the growth rate $\mathrm{Im}(\omega)$ and frequency $\mathrm{Re}(\omega)$ in top and bottom panels, respectively. Each marker represents the tearing mode in a \texttt{Legolas} run at $501$ grid points and a Laplace distributed grid was used to focus the grid points around the region of strongest change in equilibrium magnetic field at $x = 0$. Note that the non-zero $\mathrm{Re}(\omega)$ values are due to the inclusion of the Hall terms, which results in spectrum asymmetry with respect to the imaginary axis here, similar to equilibrium flow. For any guide field value $B_\mathrm{g}$, the growth rate is influenced by the wavevector, with the maximum growth rate depending on both wavevector components, $k_2$ and $k_3$. Whilst the real part of the frequency $\mathrm{Re}(\omega)$ has an extremum as a function of $k_3$ in the presence of a guide field ($B_\mathrm{g} \neq 0$), $|\mathrm{Re}(\omega)|$ increases linearly with increasing $k_3$ in the absence of a guide field ($B_\mathrm{g} = 0$), until the tearing mode is fully damped.

\begin{figure}
\centering
\includegraphics[width=\textwidth]{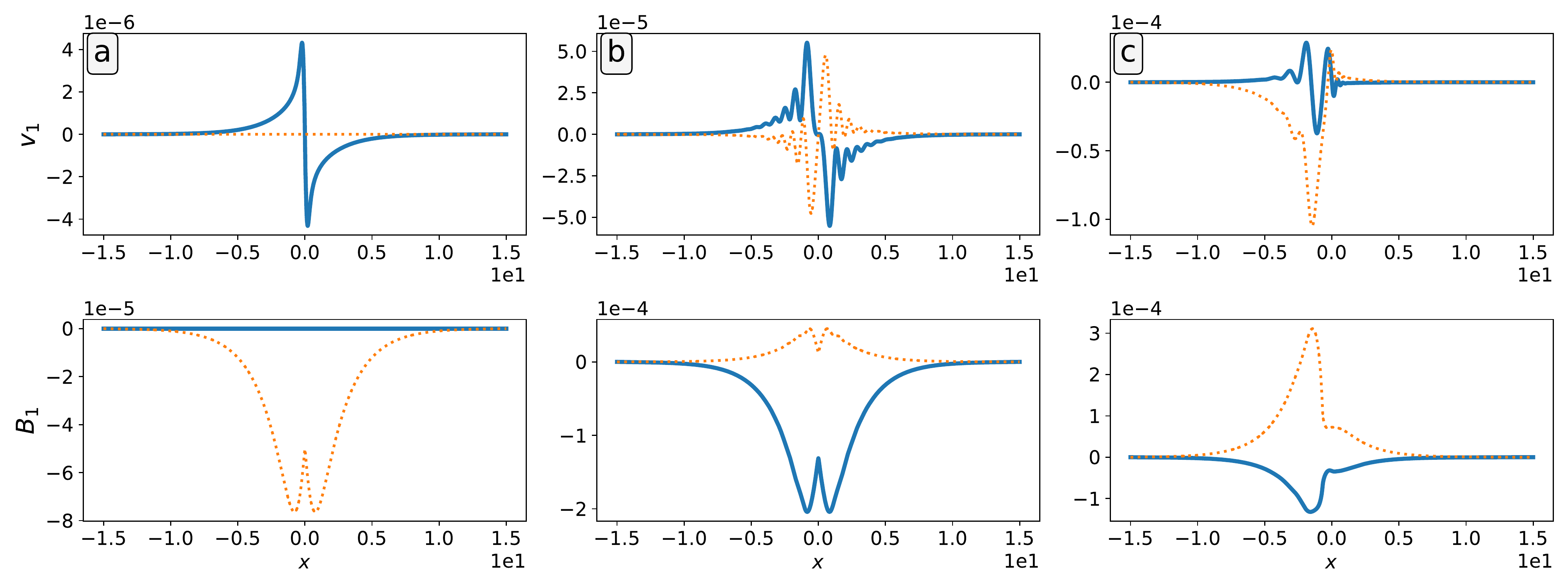}
\caption{The incompressible tearing mode's $v_1$ and $B_1$ eigenfunctions for $\eta = 10^{-4}$, $\eta_\mathrm{H} = 1$, and $k_2 = 0.5$, with different values of $k_3$ and $B_\mathrm{g}$: column (a) $k_3 = 0$, $B_\mathrm{g} = 0$; column (b) $k_3 = 0.06$, $B_\mathrm{g} = 0$; and column (c) $k_3 = 0.06$, $B_\mathrm{g} = 5$. Real parts are shown as solid (blue) lines, imaginary parts as dotted (orange) lines. All runs were performed at 501 grid points.}
\label{fig:harris-ef}
\end{figure}

Besides quantifying the tearing mode complex eigenfrequencies, \citet{Shi2020} also reported on the tearing mode eigenfunctions. Up to a complex factor, the eigenfunctions obtained by \texttt{Legolas}, shown in figure \ref{fig:harris-ef}, match the results in the first two rows of figure 7 in \citet{Shi2020}. For the cases in columns a and b of figure \ref{fig:harris-ef}, where $B_\mathrm{g} = 0$, the $B_1$ and $v_1$ eigenfunctions are symmetric and antisymmetric, respectively, with respect to the location of the Harris sheet ($x=0$) whereas the (anti)symmetry is broken with the introduction of a non-zero guide field $B_\mathrm{g}$ (column c).

\begin{figure}
\centering
\includegraphics[width=\textwidth]{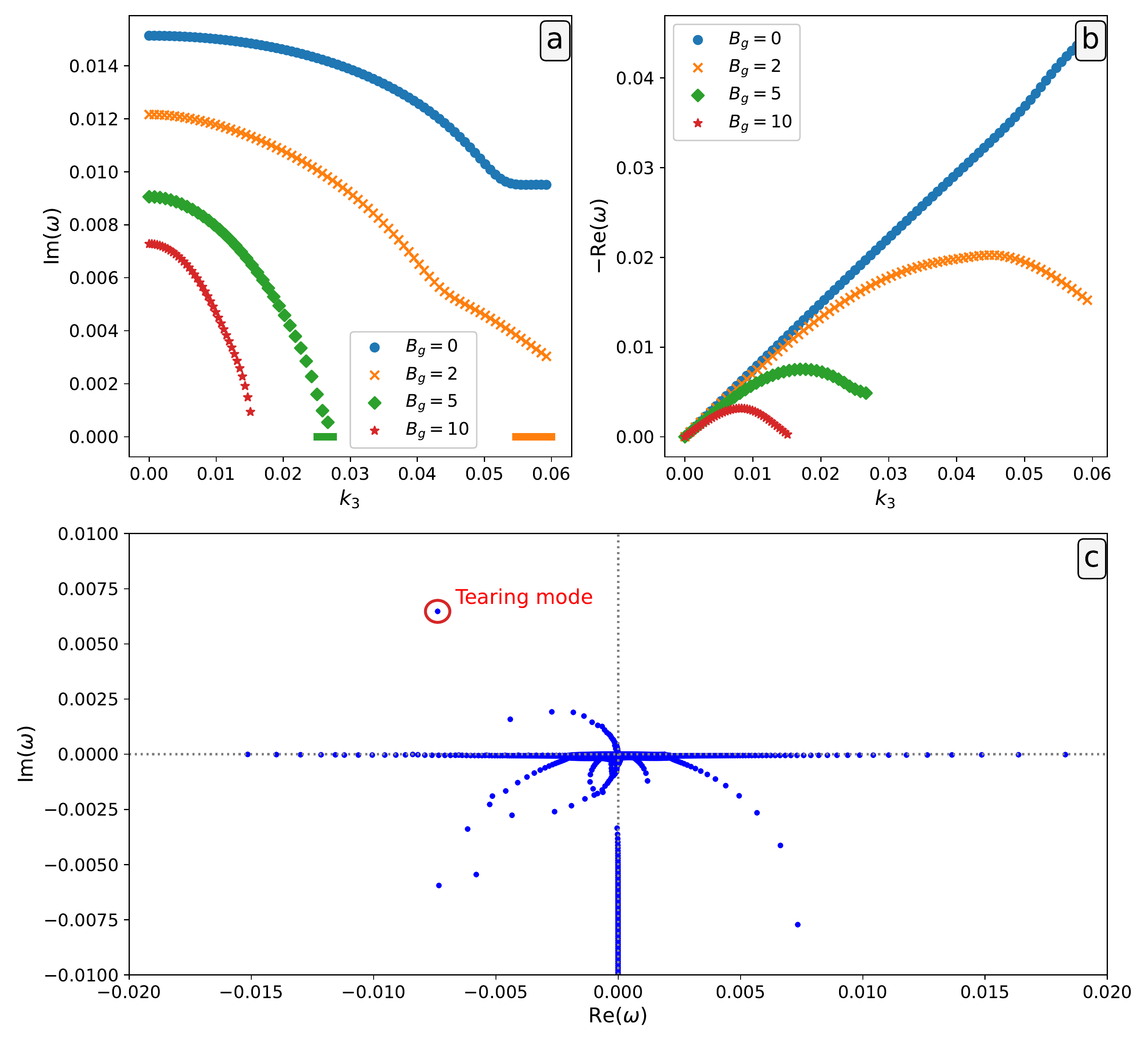}
\caption{The real (a) and imaginary (b) parts of the compressible tearing mode in a Harris current sheet (\ref{equil:hall-tearing}) as a function of $k_3$, with $k_2 = 0.155$, $\tilde{\rho}=1$, $a=1$, $B_0=1$, $\eta = 10^{-4}$, and $\eta_\mathrm{H} = 1$ for different guide field strengths $B_\mathrm{g}$, to be compared to the incompressible case in figure \ref{fig:harris}a and b. The horizontal lines in (a) indicate the ranges where the tearing mode is not the most unstable mode in the spectrum. (c) Spectrum from the $B_\mathrm{g} = 5$ series with $k_3 \approx 0.015346$. The tearing mode is circled in red. All runs were performed at $501$ grid points.}
\label{fig:compressible-diff}
\end{figure}

\citet{Shi2020} only quantify incompressible linear eigenmodes, which they justify by stating that the resistive tearing mode has a negligible contribution due to compressibility, based on the reasoning followed by \citet{Furth1963}. We can here easily verify that assumption, 
using the full compressible functionality of \texttt{Legolas}. It appears that the inclusion of Hall terms in the treatment of the tearing mode causes differences in incompressible versus compressible plasma settings. The influence of compressibility is shown in figures \ref{fig:compressible-diff}a and b, where the compressible growth rate and frequency, respectively, are shown for $k_2 = 0.155$, to be compared to the incompressible case in figures \ref{fig:harris}a and b. Although \citet{Furth1963} showed that compressibility has a negligible effect on the resistive tearing mode, which our tests with \texttt{Legolas} also confirm, their treatment did not take the Hall current into account. When the Hall terms are taken into account, the effect of compressibility on the resistive tearing mode growth rate is no longer negligible. In particular, stronger guide fields result in stronger damping of the growth rate, as evidenced by figure \ref{fig:compressible-diff}. Additionally, new unstable modes appear in the spectrum and become more unstable than the tearing mode for sufficiently large $k_3$. These are Hall instabilities, occurring in a Cartesian slab when the magnetic field is sufficiently curved, i.e. if $\partial^2\bfb_0/\partial x^2$ is non-zero \citep{Rheinhardt2002}. The ranges where the largest Hall instability overtakes the tearing instability as the most unstable mode are indicated in figure \ref{fig:compressible-diff}a with lines on the horizontal axis. The part of a spectrum containing the tearing mode and the other unstable modes is shown in figure \ref{fig:compressible-diff}c.

\section{Conclusion and outlook}\label{sec:conclusions}
In this paper, the extension of the MHD spectroscopy code \texttt{Legolas} \citep{Claes2020} with viscosity and the Hall current was presented and verified using test cases taken from the literature. To validate the implementation of the viscosity module, we first accurately reproduced the spectrum and eigenfunctions of an incompressible, hydrodynamic Taylor-Couette flow in a cylindrical setup, taken from \citet{Gebhardt1993}, with the newly implemented incompressible approximation. As a second test case, we considered the Cartesian magnetohydrodynamic equilibrium with finite resistivity from \citet{Dahlburg1983}, where we reproduced their results concerning the interplay of viscous and resistive effects on the growth rate of the resistive tearing instability. As an extension of their results, we showed that the full resistive and viscous spectrum are extremely similar, with the prominent distinction that the viscous spectrum does not have an unstable tearing mode. The combination of viscous and resistive effects was mostly seen in the stable part of the spectrum, and its role in further non-linear evolutions warrants further exploration. However, since both viscosity test cases used the incompressible approximation, which eliminates the energy equation, the viscous heating term did not play a role. For one selected Taylor-Couette case it was shown that the viscous heating did not significantly alter the spectrum, but had a limited influence on the entropy perturbation. The flexible \texttt{Legolas} implementation allows for future linear stability studies with or without viscous heating. Such viscoresistive stability studies of magnetised Taylor-Couette setups can be very important for aiding the interpretation of dynamo experiments\citep{Willis2002,Rudiger2007}, and especially to determine when the MRI is sufficiently suppressed by viscoresistive effects \citep{Eckhardt2018} to create stable configurations.

To test the implementation of the Hall module, another two cases were considered. The simplest case considered an ideal, homogeneous, Cartesian plasma slab with a Hall current. For a small ratio of ion inertial length to plate separation this case is comparable to the infinite, homogeneous medium, described by the dispersion relation of \citet{Hameiri2005}. The solutions of the infinite medium corresponded to the first modes in several sequences of modes, as evidenced by the eigenfunctions, which is expected when going from an infinite medium to a semi-infinite medium that is bounded in one direction. Whilst the smallest sequence behaves anti-Sturmian, the larger two display Sturmian behaviour. All three wave sequences become dispersive in Hall-MHD, and the smallest and largest sequences are known as ion cyclotron and whistler modes respectively. The middle sequence fails to capture the electron cyclotron resonance because the electron mass was set to zero. If the electron inertia term is included as well, the electron cyclotron resonance is recovered.

The more advanced test case also included resistivity and evaluated Hall effects on the resistive tearing mode growth rate in a Harris sheet setup described by \citet{Shi2020}. The reproduction of these results required an incompressible approximation, but a good match between both the tearing mode and the eigenfunctions was achieved. However, contrary to the assumption of \citet{Shi2020} that compressibility has a negligible effect on the resistive tearing mode, which was shown for the purely resistive MHD case by \citet{Furth1963}, a guide field introduces a non-negligible damping effect when both compressibility and the Hall current are considered.

Our \texttt{Legolas} tool can now be used to combine and explore linear eigenmodes and full eigenspectra for cases where we have multiple effects at play, such as the influence of equilibrium flow or non-uniform equilibrium density, the combination of Hall and viscosity, and the electron inertia term. However, the effect of the latter is likely negligible in many cases because the electron inertia coefficient $\eta_\mathrm{e}$ is several orders of magnitude smaller than the Hall coefficient $\eta_\mathrm{H}$.

The inclusion of viscosity and the Hall current opens up various research avenues, such as the investigation of the influence of viscosity in resistive setups, and in particular how the introduction of viscosity affects resistive instabilities like the resistive tearing mode discussed in section \ref{sec:viscoresistive}. For the Hall current it is now possible to examine its effect on the previously-mentioned MRI in accretion discs \citep{Lesur2021} or to explore instabilities requiring a Hall current, such as the Hall-shear instability \citep{Kunz2008}. In the context of the MRI, a similar future extension of the \texttt{Legolas} code can implement ambipolar diffusion as a proxy for charge-neutral decoupling effects. This would also introduce the ambipolar-diffusion-shear instability \citep{Kunz2008}.

Looking ahead, this extension brings \texttt{Legolas} one step closer to describing realistic 1D setups. We foresee that the tool can even be used to identify the modes responsible for specific evolutions seen in 2D and 3D non-linear simulations, when we can describe the instantaneous multidimensional MHD fields with a (e.g. height and azimuthally averaged) force-balanced 1D background state, and which physical effects play a relevant role in observed growth rates. Furthermore, since \texttt{Legolas} captures both fundamental modes and overtones, it can act as a tool in the analysis of observed waves and overtones in coronal loop seismology \citep[see e.g.][]{Andries2009}, albeit in the infinite cylinder approximation, and tokamaks \citep[see e.g.][]{Ochoukov2018,Spong2018}, although toroidal effects are lost in the cylinder setup. In the future, vacuum or vacuum-wall boundary conditions could be implemented to better model these physical systems.\\

\textbf{Supplementary material.} The \texttt{Legolas} code is freely available under the GNU General Public License. For more information, visit \url{https://legolas.science/}.

\textbf{Funding.} This work is supported by funding from the European Research Council (ERC) under the European Unions Horizon 2020 research and innovation programme, Grant agreement No. 833251 PROMINENT ERC-ADG 2018.

\textbf{Declaration of Interests.} The authors report no conflict of interest.

\appendix
\section{Linearised equations}\label{app:matrix}
As explained originally in \citet{Claes2020}, after the Fourier analysis (\ref{eq:fourier}) the variables are transformed as
\begin{equation}\label{eq:coordtransform}
\begin{aligned}
    &\eps\rho_1\rightarrow\tilde{\rho}_1,\quad \im\eps v_1\rightarrow \tilde{v}_1,\quad v_2\rightarrow \tilde{v}_2,\quad \eps v_3\rightarrow \tilde{v}_3, \\
    &\eps T_1\rightarrow \tilde{T}_1,\quad \im A_1\rightarrow \tilde{a}_1,\quad \eps A_2\rightarrow \tilde{a}_2,\quad A_3\rightarrow \tilde{a}_3,
\end{aligned}
\end{equation}
where $\eps$ is a scale parameter equal to $1$ in Cartesian setups and equal to $r$ in the cylindrical case. In these new variables (dropping tildes below for notational convenience), the full equations, including viscosity and Hall, are implemented in \texttt{Legolas} in the form
\begin{gather}
    \omega\rho_1 = -\rho_0' v_1 - \rho_0\Bigl(v_1' - k_2 v_2 - k_3 v_3\Bigr) + \rho_1 \left(\frac{k_2}{\eps} v_{02} + k_3 v_{03} \right), \label{eq:app-continuity} \\
    \begin{aligned}
			\omega\rho_0 v_1  = &\eps\left(\frac{\rho_1 T_0 + \rho_0 T_1}{\eps}\right)' + g \rho_1
			+ B_{02} \left\{- \frac{k_2 k_3}{\eps}a_2 + \frac{k_2^2}{\eps}a_3 + \left[\eps\left(k_3a_1 - a_3'\right)\right]'\right\} \\
			&+ B_{03} \left\{-k_3^2 a_2 + k_2 k_3a_3 + \eps\left[\frac{1}{\eps}\left(a_2' - k_2 a_1 \right)\right]'\right\}
			+ B_{03}' \left(a_2' - k_2a_1\right) \\
			&+ \left(\eps B_{02}\right)'\left(k_3a_1 - a_3'\right) - \frac{\eps'}{\eps} v_{02}^2\rho_1
			+ \rho_0  \left(\frac{k_2}{\eps} v_{02} + k_3 v_{03}\right)v_1 - 2 \eps' \rho_0 v_{02} v_2 \\
			&- \im\mu \left(\frac{\eps'}{\eps^2} + \frac{k_2^2}{\eps^2} + k_3^2\right) v_1
			+ \frac{\im\mu}{3} \eps\left(\frac{1}{\eps}v_1' - \frac{k_2}{\eps} v_2 - \frac{k_3}{\eps} v_3 \right)' \\
			&+ 2\frac{\im\mu \eps'}{\eps} k_2 v_2
			+ \im\mu \left[\eps \left(\frac{v_1}{\eps}\right)'\right]',
	\end{aligned} \label{eq:app-momentum1} \\
	\begin{aligned}
	       \omega\rho_0 \eps v_2 = &\frac{k_2}{\eps} (\rho_1 T_0 + \rho_0 T_1) + B_{03} \left[- \left(\frac{k_2^2}{\eps} + \eps k_3^2\right)a_1  + \frac{ k_2}{\eps}a_2' +  \eps k_3a_3'\right] \\
	       &+ \frac{(\eps B_{02})'}{\eps}\Bigl(k_3 a_2 - k_2 a_3\Bigr) - \frac{\left(\eps v_{02}\right)'}{\eps}\rho_0 v_1 + \rho_0 \Bigl(k_2 v_{02} + \eps k_3 v_{03}\Bigr)v_2 \\
	       &+ \frac{\im\mu}{\eps}\left(2\frac{\eps' }{\eps} k_2 v_1 + \frac{1}{3} k_2 v_1' - \frac{1}{3} k_2 k_3 v_3 \right) \\
	       &- \im\mu \left(\frac{\eps'}{\eps} + \frac{4}{3}\frac{k_2^2}{\eps} + \eps k_3^2\right) v_2 + \im\mu \left(\eps v_2'\right)',
	\end{aligned} \label{eq:app-momentum2} \\
	\begin{aligned}
	       \omega \rho_0 v_3 = &k_3 (\rho_1 T_0 + \rho_0 T_1) + B_{02} \left[\left(\frac{k_2^2}{\eps} + \eps k_3^2\right) a_1  - \frac{k_2}{\eps}a_2'  - \eps k_3 a_3' \right] \\
	       &+ B_{03}' \Bigl(k_3 a_2  - k_2 a_3 \Bigr) - \rho_0 v_{03}'v_1 + \rho_0 \left(\frac{k_2}{\eps} v_{02} + k_3 v_{03}\right) v_3 \\
	       &+ \frac{\im\mu}{3} k_3 \Bigl(v_1' - k_2 v_2\Bigr) - \im\mu \left(\frac{k_2^2}{\eps^2} + \frac{4}{3} k_3^2 \right) v_3 + \im\mu \left[\eps \left(\frac{v_3}{\eps}\right)'\right]',
	\end{aligned} \label{eq:app-momentum3} \\
	\begin{aligned}
	       \omega \bigg\{ \eps a_1 &+ \eta_\mathrm{H} v_1 + \frac{\eta_\mathrm{e}}{\rho_0} \left[ \left( \frac{k_2^2}{\eps} + \eps k_3^2 \right) a_1 - \frac{k_2}{\eps} a_2' - \eps k_3 a_3' \right] \bigg\} \\
	       = &B_{02}v_3 - \eps B_{03}v_2 + \left(k_2 v_{02} + \eps k_3 v_{03}\right)a_1  - v_{02}a_2' - \eps v_{03}a_3' \\
	       &- \im \eta_0 \left(\frac{k_2^2}{\eps} + \eps k_3^2\right)a_1 + \im \eta_0 \frac{k_2}{\eps} a_2' + \im \eta_0 \eps k_3 a_3' \\
	       &+ \eta_\mathrm{H} \left[ \left( \frac{k_2}{\eps} v_{02} + k_3 v_{03} \right) v_1 - 2\eps' v_{02} v_2 + \frac{1-f_\mathrm{e}}{\rho_0} \left( \rho_0' T_1 - T_0' \rho_1 \right) \right] \\
	       &+\im\mu\frac{\eta_\mathrm{H}}{\rho_0} \bigg\{ \left[ \eps \left( \frac{v_1}{\eps} \right)' \right]' - \left( \frac{k_2^2}{\eps^2} + k_3^2 \right) v_1 + 2\frac{\eps'}{\eps} k_2 v_2 \\
	       &\hspace{2cm} - \frac{\eps'}{\eps^2} v_1 + \frac{\eps}{3} \left[ \frac{1}{\eps} \left( v_1' - k_2 v_2 - k_3 v_3 \right) \right]' \bigg\},
	\end{aligned} \label{eq:app-induction1} \\
	\begin{aligned}
	       \omega \Bigg\{ a_2 &+ \eta_\mathrm{H}\eps v_2 + \frac{\eta_\mathrm{e}}{\rho_0} \left[ \eps \left( \frac{1}{\eps} (k_2 a_1 - a_2') \right)' + k_3 (k_3 a_2 - k_2 a_3) \right] \Bigg\} \\
	       = &-B_{03} v_1 + v_{03} (k_3 a_2 - k_2 a_3) + \im B_{03}' \frac{\mathrm{d}\eta}{\mathrm{d}T} T_1 - \im\eta_0 k_3 (k_3 a_2 - k_2 a_3) \\
	       &+ \im\eta_0\eps \left(\frac{1}{\eps}a_2' - \frac{k_2}{\eps}a_1\right)' + \eta_\mathrm{H} \left[ \left( k_2 v_{02} + \eps k_3 v_{03} \right) v_2 - \left(v_{02}' - \frac{\eps'}{\eps} v_{02} \right) v_1 \right] \\
	       &+ \im\mu\frac{\eta_\mathrm{H}}{\rho_0} \bigg\{ \left( \eps v_2' \right)' - \left( \frac{k_2^2}{\eps} + \eps k_3^2 \right) v_2 + 2\frac{\eps'}{\eps^2} k_2 v_1 - \frac{\eps'}{\eps} v_2 \\
	       &\hspace{2cm} + \frac{1}{3} \frac{k_2}{\eps} \left( v_1' - k_2 v_2 - k_3 v_3 \right) - \frac{\rho_1}{\rho_0} \left( v_{02}'' + \frac{\eps'}{\eps} v_{02}' - \frac{\eps'}{\eps^2} v_{02} \right) \bigg\},
	\end{aligned} \label{eq:app-induction2} \\
	\begin{aligned}
	       \omega \bigg\{ \eps a_3 &+ \eta_\mathrm{H} v_3 + \frac{\eta_\mathrm{e}}{\rho_0} \left[ \left( \eps (k_3 a_1 - a_3') \right)' - \frac{k_2}{\eps} (k_3 a_2 - k_2 a_3) \right] \bigg\} \\
	       = &B_{02} v_1 - v_{02} (k_3 a_2 - k_2 a_3) - \im \frac{\left(\eps B_{02} \right)'}{\eps} \frac{\mathrm{d}\eta}{\mathrm{d}T} T_1 + \im \eta_0 \frac{k_2}{\eps} (k_3 a_2 - k_2 a_3) \\
	       &- \im \eta_0 \Bigl(\eps k_3 a_1 - \eps a_3'\Bigr)' + \eta_\mathrm{H} \left[ \left( \frac{k_2}{\eps} v_{02} + k_3 v_{03} \right) v_3 - v_{03}' v_1 \right] \\
	       &+ \im\mu\frac{\eta_\mathrm{H}}{\rho_0} \bigg\{ \left[ \eps \left( \frac{v_3}{\eps} \right)' \right]' - \left( \frac{k_2^2}{\eps^2} + k_3^2 \right) v_3 + \frac{k_3}{3} \left( v_1' - k_2 v_2 - k_3 v_3 \right) \\
	       &\hspace{2cm} - \frac{\rho_1}{\rho_0} \left( v_{03}'' + \frac{\eps'}{\eps} v_{03}' \right) \bigg\},
	\end{aligned} \label{eq:app-induction3} \\
	\begin{aligned}
	       \omega\rho_0 T_1 = &- \rho_0 T_0' v_1 + \rho_0 \left(\frac{k_2}{\eps} v_{02} + k_3 v_{03}\right) T_1 - (\gamma - 1) \rho_0 T_0 \Bigl(v_1' - k_2 v_2 - k_3 v_3\Bigr) \\
	       &- \im(\gamma - 1) \HL_T \rho_0 T_1  - \im(\gamma - 1) \Bigl(\HL_0 + \HL_\rho \rho_0\Bigr) \rho_1 \\
	       &- \im(\gamma - 1)\prefactkappazero\Foperator^2 T_1 - \im(\gamma - 1)\kappa_{\perp, 0}\left(\frac{k_2^2}{\eps^2} + k_3^2\right) T_1 \\
	       &+ \im(\gamma - 1)\prefactkappazero T_0' \Foperator\Bigl(k_3 a_2 - k_2 a_3\Bigr) + \im(\gamma - 1)\Bigl(\eps T_0' \kappa_{\perp, 1}\Bigr)' \\
	       &+ \im(\gamma - 1)\left[\eps\kappa_{\perp, 0}\left(\frac{T_1}{\eps}\right)'\right]' + \im (\gamma - 1) T_1 \frac{\mathrm{d}\eta}{\mathrm{d}T} \left[B_{03}'^2 + \left(\frac{\left(\eps B_{02}\right)'}{\eps}\right)^2\right] \\
	       &+ 2\im(\gamma - 1) \eta_0 \Bigg\{B_{03}' \left[ k_3 (k_2 a_3 - k_3 a_2) + \eps\left( \frac{1}{\eps}a_2' - \frac{k_2}{\eps}a_1 \right)'\right] \\
	       &\hspace{2.5cm} + \frac{\left(\eps B_{02}\right)'}{\eps}\left[\frac{k_2}{\eps} (k_2 a_3 - k_3 a_2) + \left(\eps k_3 a_1 - \eps a_3'\right)'\right]\Bigg\} \\
	       &+ 2\im\mu \left[ \frac{(\eps')^2}{\eps} v_{02} v_2 + \eps v_{02}' v_2' + \eps v_{03}' \left(\frac{v_3}{\eps}\right)' - \frac{\eps'}{\eps^2} k_2 v_{02} v_1 \right],
	\end{aligned} \label{eq:app-energy}
\end{gather}
where a prime denotes derivation with respect to $u_1$ and the notation
\begin{equation}
       \Foperator = \Foperatorfull
\end{equation}
was introduced.

\bibliographystyle{jpp}
\bibliography{bibliography}

\end{document}